\documentclass{article}

\usepackage{graphicx}

\begin{document}

\title{Critical Data Compression}

\author{John Scoville}

\maketitle

\begin{abstract}
A new approach to data compression is developed and applied to multimedia content.  This method separates messages into components suitable for both lossless coding and 'lossy' or statistical coding techniques, compressing complex objects by separately encoding signals and noise.  This is demonstrated by compressing the most significant bits of data exactly, since they are typically redundant and compressible, and either fitting a maximally likely noise function to the residual bits or compressing them using lossy methods.  Upon decompression, the significant bits are decoded and added to a noise function, whether sampled from a noise model or decompressed from a lossy code.  This results in compressed data similar to the original.  Signals may be separated from noisy bits by considering derivatives of complexity in a manner akin to Kolmogorov's approach or by empirical testing.  The critical point separating the two represents the level beyond which compression using exact methods becomes impractical.  Since redundant signals are compressed and stored efficiently using lossless codes, while noise is incompressible and practically indistinguishable from similar noise, such a scheme can enable high levels of compression for a wide variety of data while retaining the statistical properties of the original.  For many test images, a two-part image code using JPEG2000 for lossy compression and PAQ8l for lossless coding produces less mean-squared error than an equal length of JPEG2000.  For highly regular images, the advantage of such a scheme can be tremendous.  Computer-generated images typically compress better using this method than through direct lossy coding, as do many black and white photographs and most color photographs at sufficiently high quality levels.  Examples applying the method to audio and video coding are also demonstrated.  Since two-part codes are efficient for both periodic and chaotic data, concatenations of roughly similar objects may be encoded efficiently, which leads to improved inference.  Such codes enable complexity-based inference in data for which lossless coding performs poorly, enabling a simple but powerful minimal-description based approach audio, visual, and abstract pattern recognition.  Applications to artificial intelligence are demonstrated, showing that signals using an economical lossless code have a critical level of redundancy which leads to better description-based inference than signals which encode either insufficient data or too much detail.
\end{abstract}

\section{Complexity and Entropy}
In contrast to information-losing or 'lossy' data compression, the lossless compression of data, the central problem of information theory, was essentially opened and closed by Claude Shannon in a 1948 paper\cite{SH48}.  Shannon showed that the entropy formula (introduced earlier by Gibbs in the context of statistical mechanics) establishes a lower bound on the compression of data communicated across some channel - no algorithm can produce a code whose average codeword length is less than the Shannon information entropy.  If the probability of codeword symbol $i$ is $P_{i}$:
\begin{equation}
S = -k \sum{ P_{i} \log{P_{i}} }
\end{equation}

This quantity is the amount of information needed to invoke the axiom of choice and sample an element from a distribution or set with measure; any linear measure of choice must have its analytic form of expected log-probability\cite{SH48}.  This relies on the knowledge of a probability distribution over the possible codewords.  Without a detailed knowledge of the process producing the data, or enough data to build a histogram, the entropy may not be easy to estimate.  In many practical cases, entropy is most readily measured by using a general-purpose data compression algorithm whose output length tends toward the entropy, such as Lempel-Ziv.  When the distribution is uniform, the Shannon/Gibbs entropy reduces to the Boltzmann entropy function of classical thermodynamics; this is simply the logarithm of the number of states.

The entropy limit for data compression established by Shannon applies to the exact ('lossless') compression of any type of data.  As such, Shannon entropy corresponds more directly to written language, where each symbol is presumably equally important, than to raw numerical data, where leading digits typically have more weight than trailing digits.  In general, an infinite number of trailing decimal points must be truncated from a real number in order to obtain a finite, rational measurement.  Since some bits have much higher value than others, numerical data is naturally amenable to information-losing ('lossy') data compression techniques, and such algorithms have become routine in the digital communication of multimedia data.  For the case of a finite-precision numerical datum, rather than the Shannon entropy, a more applicable complexity measure might be Chaitin's algorithmic prefix complexity\cite{CH87} which measures the irreducible complexity of the leading digits from an infinite series of bits.  The algorithmic prefix complexity is an example of a Kolmogorov complexity\cite{LV97}, the measure of minimal descriptive complexity playing a central role in Kolmogorov's formalization of probability theory.

Prior to the twentieth century, this basic notion of a probability distribution function (pdf) had not changed significantly since the time of Gauss.  After analysis of the Brownian motion by Einstein and others, building on the earlier work of Markov, the stochastic process became a popular idea.  Stochastic processes represent the fundamental, often microscopic, actions which lead to frequencies tending, in the limit, to a probability density.  Stochastic partial differential equations (for example, the Fokker-Planck equation) generate a pdf as their solution, as do the 'master equations' from whence they are derived; such pdfs may describe, for instance, the evolution of probabilities over time.  They were used notably by Langevin to separate dynamical systems into a deterministic classical part and a random stochastic component or statistical model.  Given empirical data from such a system, the Langevin approach may be combined with the maximum likelihood method\cite{FI21} or Bayesian inference (maximum posterior method) to identify the most likely parameters for an unknown noise function.

In practice, Langevin's approach either posits the form of a noise function or fits it to data; it does not address whether or not data is stochastic in the first place.  Kolmogorov addressed this issue, refining the notion of stochastic processes and probability in general.  Some objects, a solid black image, for example, are almost entirely regular.  Other data seem totally random; for instance, geiger counters recording radioactive decays.  Stochastic objects lay between these two extremes; as such, they exhibit both deterministic and random behavior.  Kolmogorov introduced a technique for separating a message into random and nonrandom components.  First, however, he defined the Kolmogorov complexity, $C(X)$.  $C(X)$ is the minimum amount of information needed to completely reconstruct some object, represented as a binary string of symbols, X\cite{CO91,LV97}.
\begin{equation} C_{f}(X) = \min_{f(p)=X}|p| \end{equation}

The recursive function f may be regarded as a particular computer and p is a program running on that computer.  The Kolmogorov Complexity is the length of the shortest computer program that terminates with X as output on computer f.  In this way, it symbolizes perfect data compression.  For various reasons (such as the non-halting of certain programs), it is usually impossible to prove that non-trivial representations are minimal.  On the other hand, a halting program always exists, the original string, so a minimal halting program also exists, even if its identity can't be verified.  In practice, the Kolmogorov complexity asymptotically approaches the Shannon entropy\cite{LV97}, and the complexity of typical objects may be readily approximated using the length of an entropic code.

Often, a variant of Kolmogorov complexity is used - Chaitin's algorithmic prefix complexity $K(X)$\cite{CH87} which considers only self-delimiting programs that do not use stop symbols.  Since a program may be self-delimiting by iteratively prefixing code lengths, $K(X) = C(X) + C(C(X)) + O(C(C(C(X))))$\cite{LV97}

Returning to the separation of signal and noise, we now define stochasticity as it relates to the Kolmogorov structure function.  For natural numbers $k$ and $\delta$, we say that a string x is $(k,\delta)$-stochastic if and only if there exists a finite set $A$ such that:
\begin{equation}
x \in A, C(A) \leq k, C(x|A)\geq log |A| - \delta
\end{equation}

The deviation from randomness, $\delta$, indicates whether x is a typical or atypical member of A.  This is minimized through the Kolmogorov Structure Function, $C_{k}(x|n)$:
\begin{equation}
C_{k}(x|n) = \min {\{\log{|A|}: x \in A,C(A|n) \leq k\}}
\end{equation}

The minimal set $A_{0}$ minimizes the deviation from randomness, $\delta$, and is referred to as the Kolmogorov Minimal Sufficient Statistic for x given n.  The Kolmogorov Structure Function specifies the bits of additional entropy (Shannon entropy reduces to the logarithm function for a uniform distribution) necessary to select the element x from a set described with k or fewer bits.  For a regular object, the structure function has a slope less than -1.  Specifying another bit of k reduces the entropy requirement by more than a bit, resulting in compression.  Beyond a critical threshold, corresponding to the minimal sufficient statistic, stochastic objects become random.  Beyond this point, specifying another bit of k increases the entropy by exactly one bit, so the slope of the structure function reaches its maximum value of -1.  For this reason, Kolmogorov identified the point at which the slope reaches -1 as the minimal sufficient statistic.  The Kolmogorov minimal sufficient statistic represents the amount of information needed to capture all the regular patterns in the string x without literally specifying the value of random noise.

While conceptually appealing, there are practical obstacles to calculating the Kolmogorov minimal sufficient statistic.  First, since the Kolmogorov complexity is not directly calculable, neither is this statistic. Approximations may be made, however, and when using certain common data compression algorithms, the point having slope 1 is actually a reasonable estimate of the onset of noise.  When certain data are compressed more fully, however, this point may not exist.  For example, consider a color photograph of black and white static on an analog TV set.  The pattern of visible pixels emerges from nearly incompressible entropy; chaos resulting from the machine's attempt to choose values from a nonexistent signal.  Since a color photograph has three channels, and the static is essentially monochromatic; the channels are correlated to one another and hence contain compressible mutual information.  As such, the noise in the color photograph, though emergent from essentially pure entropy, is intrinsically compressible - hence, the compression ratio never reaches 1:1 and the Kolmogorov minimal sufficient statistic does not exist.

Instead of the parameter value where the compression ratio reaches 1:1, which may not exist, one often seeks the parameter value which provides the most information about the object under consideration.  The problem of determining the most informative parameters in a model was famously addressed by the statistician R.A. Fisher\cite{FI21}.

The Fisher Information quantifies the amount of information expected to be inferred in a local neighborhood of a continuously parameterizable probability distribution.  The Fisher Information quantifies information at specific values of the parameters - it quantifies the informative-ness of a local observation.

If the probability density of X is parameterized along some path by t, f(X;t), then the Fisher Information Metric at some value of t is the expectation of the variance of the Hartley information\cite{HA28}, also known as the score:
\begin{equation}
I(t)=-E[(\partial_t \ln{f(X;t)})^2 | t]
\end{equation}

The Fisher Information quantifies the convexity (the curvature) of an entropy function at a specific point in parameter space, provided sufficient regularity and differentiability.

In the case of multiple parameters, the Fisher Information becomes the Fisher Information Metric (or Fisher Information Matrix, FIM), the expected covariance of the score:
\begin{equation}
I(t)_{ij} = E[\partial_{t_i} \ln{f(X;t)} \partial_{t_j} \ln{f(X;t)}]
\end{equation}

The Fisher-Rao metric is simply the average of the metric implied by Hartley information over a parameterized path.  The space described by this metric has distances that represent differences in information or entropy.  The differential geometry of this metric is sometimes called \emph{information geometry}.  We seek the parameter values maximizing the Fisher-Rao metric, for variations in these values lead to the largest possible motions in the metric space of information.

If we take $f(X;t)$ to be the universal probability of obtaining a string X from a randomly generated program on a universal computer, this probability is typically dominated by the shortest possible program, implying that $f(X;t) \approx 2^{-K(X)}$, where we have used $K(X)$ instead of $C(X)$ so the sum over all X will convergence to unit probility\cite{LV97}.  If we make the string $X$ a function of some parameter t, $X=X(t)$, then $f(X;t) \approx 2^{-K(X(t))}$, the Hartley information is $\ln 2$ times $-K(X(t))$, and its associated Fisher-Rao metric is:
\begin{equation}
I(t)_{ij} \approx  (\ln 2)^2 E[ \partial_{t_i} K(X(t)) \partial_{t_j} K(X(t)) ]
\end{equation}

Since the spaces we consider are generally discrete, we will consider paths from one parameter value to the next and evaluate partial differences in place of the partial derivatives.  The one-dimensional Fisher information of a path from $n$ to $n+1$ is, replacing the continuous differentials with finite differences, and ignoring the expectation operator, which becomes the identity operator since the expectation covers only one point:
\begin{equation}
I(t) = 2 \ln 2 (K(X(t+1)-K(X(t)))^2, 0<n<d
\end{equation}

Maximizing this quantity is equivalent to maximizing $K(X(t+1)-K(X(t))$, which is also the denominator in the slope of the Kolmogorov structure function.  For incompressible data, the numerator $\log{|A_{t}|}-\log{|A_{t-1}|}$ (which is the number of additional bits erased by the uncompressed $X(t)$ beyond those erased by the more descriptive $X(t+1)$) also takes on this value.  Since the parameter in the Kolmogorov structure function corresponds to bits of description length, the literal description corresponding to each subsequent parameter value differs in length by a constant, minimizing the slope of the Kolmogorov structure function is equivalent to maximizing $K(X(t+1)-K(X(t))$ and the Fisher information.  The minimal parameter that maximizes the Fisher information is the Kolmogorov minimal sufficient statistic.

Sometimes, rather than considering the point at which a phase transition is complete, we wish to consider the critical point at which it proceeds most rapidly.  For this, we use the expectation of the Hessian of the Hartley information:
\begin{equation}
J(t)_{ij} = E[\partial^2_{t_i} \ln{f(X;t)} \partial^2_{t_j} \ln{f(X;t)}]
\end{equation}
This is in contrast to the expectation of the Hartley information, the entropy, or the expected curvature of the Hartley information, the Fisher information.  When this function is maximized, the Fisher information (or slope of the Kolmogorov structure function) is changing as rapidly as possible.  This means that the phase transition of interest is at its critical point and proceeding at its maximum rate.  Beyond this point, the marginal utility of each additional bit decreases as the phase transition proceeds past criticality to completion at the minimal sufficient statistic.

The derivatives in the Fisher information were approximated using backwards differences in complexity; however, a forward second difference may applied subsequently to complete the Hessian, the net result of this is a central difference approximation to the second derivative of complexity: $\partial^2_{t_i} K(A^{t_i}) \approx K(A^{t+1}) - 2K(A^{t}) + K(A^{t-1})$.  The maximum resulting from this approximation is between the maximum and minimum values of the parameter, exclusively.

In practice, since we can't calculate $K(X)$ exactly, it is helpful to treat any value of the Fisher information (or the slope of the Kolmogorov structure function) within some tolerance of the maximum $K(X^n)-K(X^{n-1})$ as being a member of a nearly-maximum set, and select the element of this set having the fewest bits.  Usually, the representation having the lowest complexity is the one with the lowest bit depth or resolution, but not always - when lossless compression is applied to highly regular objects, the lossless representation may be simpler than any 2-part or purely lossy code.  This statistic the represents all the bits of signal that can be described before additional bits require a nearly maximal description - it quantifies the minimum complexity needed to complete a phase transition from a low-complexity periodic signal to a high-complexity chaotic one.  This also applies to the maximum of the second derivative, as considered above.  Any value of the Hessian that is within some tolerance of the maximal $K(A^{t+1}) - 2K(A^{t}) + K(A^{t-1})$ is considered part of a nearly-maximal set, and the simplest element of this set is selected as the critical point.

The sufficiency of a statistic was also defined by Fisher in 1921\cite{FI21}.  If a statistic is sufficient, then no other statistic provides any additional information about the underlying distribution.  Fisher also demonstrated the relationship between maximum likelihood and sufficient statistics.  The Fisher-Neyman factorization theorem says that for a sufficient statistic T(x), the probability density $f_{\theta} (x)$ factors into terms dependent and independent of the parameter: $f_{\theta} (x) = h(x)g_{\theta}(T(x))$.  The maximum likelihood function for the parameter $\theta$ depends only the sufficient statistic $T(x)$.  As a result, a sufficient statistic is ideal for determining the parameters of a distribution using the popular method of maximum likelihood estimation (MLE).  The most efficient possible articulation of a sufficient statistic is a minimal sufficient statistic.  A sufficient statistic $S(x)$ is minimal if and only if, for all sufficient statistics $T(x)$, there exists a function f such that $S=f(T(x))$.

Partitioning the information content of a string into the complexity of its signal and the entropy of its noise is a nuanced idea that takes on several important forms, another is the algorithmic entropy\cite{ZU89,LV97} $H(Z)$ of a string.  This is defined in its most basic form as:
\begin{equation}
H(Z) = K(Z) + S
\end{equation}

In this context, $Z=X_{1:m}$ is a description of a macroscopic observation constructed by truncating a microscopic state X to a bit string of length m. K(Z) is the algorithmic prefix complexity\cite{CH87,LV97} of this representation, and $S=\log_2 n$ is the Boltzmann entropy divided by its usual mulplicative constant, k, the Boltzmann constant, and $ln 2$, since we are using bits.  n is the logarithm of the multiplicity or volume of truncated states, having universal recursive measure $2^{-m}$, so $S=-m$ and the algorithmic entropy is $H(Z)=K(Z)-m$.  This function is also known as the Martin L\"{o}f universal randomness test and plays a central role in the theory of random numbers.

The partitioning of a string into signal and noise also allows the determination of the limit to its lossy compression\cite{me1}, relative to a particular observer.  If P(X) is the set of strings which some macroscopic observer P cannot distinguish from string X, then the simplest string from this set is the minimal description equivalent to X: \begin{equation} S_{f}(X/P)
\equiv \min_{f(p) \in P(X)}{|p|} \end{equation}

We refer to this complexity as the macrostate complexity\cite{me1} or macrocomplexity since its criterion of indistinguishability corresponds to the definition of a macrostate in classical thermodynamics; a macrostate is a set of indistinguishable microstates.  Likewise, its entropy function has the form (logarithm of cardinality) of the Boltzmann entropy; it may be shown that if the probability $p(X)$ of the class $P$ is dominated by the shortest program in the class such that $p(X) \approx 2^{-K(X)}$, the macrocomplexity $S(X/P)$ is approximately:
\begin{equation} K(X) \approx S(X/P)+ \log{|X/P|} \end{equation}

This first-order approximation to macrocomplexity is close to the effective complexity of Gell-Mann and Lloyd\cite{GL96}.  The effective complexity, Y, is summed with the Shannon entropy, I, or an even more general entropy measure, such as R\'{e}nyi entropy, to define an information measure, the total information $\Sigma= Y + I$, that is typically within a few bits of K(X)\cite{GL96}.

\section{Critical Data Compression}

\subsection{Qualitative Discussion}

Critical data compression codes the most significant bits of an array of data losslessly, since they are typically redundant, and either fits a statistical model to the remaining bits or compresses them using lossy data compression techniques.  Upon decompression, the significant bits are decoded and added to a noise function which may be either sampled from a noise model or decompressed from a lossy code.  This results in a representation of data similar to the original.  This type of scheme is well-suited for the representation of noisy or stochastic data.

Attempting to find short representations of the specific states of a system which has high entropy or randomness is generally futile, as chaotic data is incompressible.  As a result, any operation significantly reducing the size of chaotic data must discard information, and this is why such a process is colloquially referred to as 'lossy' data compression.

Today, lossy compression is conventionally accomplished by optionally preprocessing and/or partitioning data and then decomposing data blocks onto basis functions.  This procedure, canonicalized by the Fourier transform, is generally accomplished by an inner product transformation projecting the signal vector onto a set of basis vectors.  However, this is not an appropriate mathematical operation for stochastic data.  Stochastic variables are not generally square-integrable, meaning that their inner products do not technically exist.  Though a discrete Fourier transform may be applied to a stochastic time series sampled at some frequency, the resulting spectrum of the sample will not generally be the spectrum of the process, as Parseval's theorem need not apply in the absence of square-integrability.

Worse, Fourier transforms such as the discrete cosine transform do not properly describe the behavior of light emitted from complex geometries.  A photograph is literally a graph of a cross-section of a solution to Maxwell's equations.  The first photographs were created by the absorption of photons on silver chloride surface, for instance.  While it is true that the solution to Maxwell's equations in a vacuum take the form of sinusoidal waves propagating in free space, a photograph of a vacuum would not generally be very interesting and, furthermore, the resolution of macroscopic photographic devices is nowhere close to the sampling frequency needed to resolve individual waves of visible light, which typically have wavelengths of a few hundred nanometers.  In a limited number of circumstances, this is appropriate - for example, a discrete cosine transformation would be ideal to encode photons emerging from a diffraction grating with well-defined spatial frequencies.  In general, however, most photographs are sampled well below the Nyquist rate necessary to reconstruct the underlying signal, meaning that microscopic detail is being lost to the resolution of the optical device used.

If a photographic scene contains multiple electrons or other charged particles, the resulting wavefront will no longer be sinusoidal, instead being a function of the geometry of charges.  Though the sine and cosine functions are orthogonal, they are complete in the sense that they may be used as a basis to express any other function.  However, since sinusoids do not generally solve Maxwell's equations in the presence of boundary conditions, the coefficients of such an expansion do not correspond to the true modes that carry energy through the electromagnetic field.  The correct set of normal modes - which solve Maxwell's equations and encode the resulting light - will be eigenfunctions or Green's functions of the geometry and acceleration of charges\cite{Jackson}.  For example, when designing waveguides (for example, fiber optics) the choice of a circular or rectangular cross-section is crucially important as this geometry determines whether the electric or the magnetic field is allowed to propagate along its transverse dimension.  Calculating the Fourier cosine transform of these fields produces a noisy spectrum; however, expanding over transverse electric and magnetic modes could produce an idealized spectrum that has all of its intensity focused into a single mode and no amplitude over the other modes.  The proper, clean spectrum is appropriate for information-losing approximations - since the (infinite) spectrum contains no energy beyond the modes under consideration, it can be truncated without compromising accuracy.  For the complex electronic geometries and motions that comprise interesting real-world photograph, these modes may be difficult to calculate, but they still exist as solutions of Maxwell's equations.  Attempting to describe them using sinusoids that don't solve Maxwell's equations leads to incompressible noise and spectra that can't be approximated accurately.

For audio, however, we note that the situation is somewhat different.  Audio signals have much lower frequency than visible light, so they are sampled above the Nyquist rate.  44,100Hz is a typical sampling rate which faithfully reconstructs sinusoids having frequency components less than 22,050Hz, which includes the vast majority of audible frequencies.  Auditory neurons will phase-lock directly to sinusoidal stimuli, making audio perception amenable to Fourier-domain signal processing.  If compressed in the time domain, the leading bits (which exhibit large, rapid oscillations) often appear more random to resource-limited data compressors than the leading bits of a Fourier spectrum.  At the same time, the less-important trailing bits are often redundant, given these leading bits, owing to vibrational modes which vary slowly compared to the sampling rate.  This reverses the trend observed in most images - their most significant bits are usually smoother and more redundant than trailing bits.  In the strictly positive domain of Fourier-transformed audio, however, the leading bits become smoother, due the use of an appropriate sampling rate.  For macroscopic photographic content, however, individual waves cannot be resolved, making Fourier-domain optical processing less effective.

Nonetheless, scientists and engineers in all disciplines all over the world successfully calculate Fourier transforms of all sorts of noisy data, and a large fraction (if not the vast majority) of all communication bandwidth is devoted to their transmission.  JPEG images use discrete cosine transforms, a form of discrete Fourier transform, as does MP3 audio and most video codecs.  Other general-purpose transformations, such as wavelets, are closely related to the Fourier transform and still suffer from the basic problem of projecting stochastic data onto a basis - the inner products don't technically exist, resulting in a noisy spectrum.  Furthermore, since the basis used doesn't solve generally solve Maxwell's equations, finite-order approximations that truncate the spectrum will not translate into results that are accurate to the desired order.  As such, we seek an alternative to expanding functions over a generic basis set.

\subsection{Compressing Data with Two-Part Codes}

The limit of lossy data compression is the Kolmogorov complexity of the macrostate perceived by the observer\cite{me1}.  Explicitly describing the macrostates perceptually coded by a human observer is prohibitively difficult, which makes optimal compression for a human observer intractable, even if the complexity were exactly calculable.  However, the truncation of data which appears in the definition of prefix complexity provides a very natural means of separating 2-part codes, the prefix complexity appears in the definition of the algorithmic entropy\cite{ZU89}, which is a special case of macrostate complexity.  Truncation of amplitude data provides a simple but universal model of data observation - an observer should regard the most significant bits of a datum as being more important than its least significant bits.

%The depth at which significance ends depends on the algorithmic entropy of the data and our objectives for the resulting representation.  This provides a simple way to separate images into significant parts, signals, and insignificant parts, noise.

%One slight difficulty in applying algorithmic entropy directly to audio, video, and image data is that such data objects are not typically ordered from their most significant to least significant bits.  As such, the algorithmic entropy as originally defined\cite{ZU89} would not correspond to effective lossy data compression of such an ordering on any reasonable device assigning greater amplitudes to more significant bits.  We desire a macrocomplexity that may be transformed into algorithmic entropy, which has a simpler definition, and this may be obtained simply by sorting the bits by significance.  The transformation between string A and string B, containing samples of bit depth d, may typically be expressed as a low-complexity permutation $A_i = B_{id+i\ mod\ d}$.  In this case, the algorithmic entropy and macrocomplexity agree, differing by at most the complexity of this permutation, which is O(1).

The codes described in this paper are the sum of a truncated macrostate, $Z=X_{1:m}$, which we call the signal, as well as a lossy approximation of the bits that were truncated from this signal, which we will refer to as that signal's residual noise function.  This is in contrast to the algorithmic entropy, which combines a truncated macrostate with all the information needed to recover its microstate.  If $n$ samples are truncated from $d$ bits to $m$, the Boltzmann entropy is proportional to $S \sim \log 2^{n(d-m)}=n(d-m)$ and the algorithmic entropy is $H(Z)=K(Z)+S=K(Z)+n(d-m)$, however, since we only store $K(Z)$ bits using lossless compression, the savings resulting from a two-part code (compared to a lossless entropic code) could approach the Boltzmann entropy $n(d-m)$ in the case of a highly compressed lossy representation.

First, the bits of datum $Y$ are reordered in the string $X$, from most significant to least significant.  This simplifies truncation and its correspondence to the conditional prefix complexity.  The resulting string is truncated to various depths, and the compressibility of the resulting string is evaluated.  The point beyond which the object has attains maximum incompressibility also maximizes the Fisher information associated with the distribution $P(X)=2^{-K(X)}$.  As discussed in the previous section, this phase transition proceeds at its maximum rate when the expected Hessian of the Hartley information, $J(t)$, is maximal.

Since the phase transition between periodicity and chaos is generally somewhat gradual, several possible signal depths could be used, to varying effect.  Following Ehrenfest's categorization of critical points by the derivative which is discontinuous, we will also refer to critical points by the order of derivatives.  Due to the discrete nature of our analysis, our difference approximations never become infinite, instead, we seek the maxima of various derivatives of the information function.

In the first-order approximation to the universal probability, the Hartley information is simply the complexity $\log P(X) = -K(X)$, therefore, multiplying the problem by -1, we classify critical points by the derivatives of complexity which have minima there.  The first of these, $I(t)=\partial_{t_i} K(X) \partial_{t_j} K(X)$, corresponds to the Fisher-Rao metric, and its maxima correspond to sufficient statistics.  If the object is chaotic beyond some level of description, then this level is also the Kolmologorov minimal sufficient statistic.  The second order matrix, $J(t) = \partial^2_{t_i} K(X) \partial^2_{t_j} K(X)$, is the point at which the Fisher Information increases most rapidly and hence the point beyond which descriptional complexity results in diminishing returns.   Higher-order critical points may be considered as well, but become progressively more difficult to determine reliably in the presence of imperfect complexity estimates, so we will analyze only the first two orders.

The choice of a first order critical point ($\max I(t)$) or a second order critical point ($\max J(t)$) as a cutoff for data compression will reflect a preference for fidelity or economy, respectively.  Other considerations may lead to alternative signal depths - the mean squared errors of images having various cutoffs are considered in the examples section, for instance.  Regardless of the critical point chosen, the redundant, compressible signal component defined by the selected cutoff point is isolated and compressed using a lossless code.

Ideally, an accurate statistical model of the underlying phenomenon, possibly incorporating psychological or other factors, would be fit to the noise component using maximum likelihood estimation to determine the most likely values of the parameters of its distribution.  Instead of literally storing incompressible noise, the parameters of the statistical model are stored.  When the code is decompressed, the lossless signal is decompressed, while the noise is simulated by sampling from the distribution of the statistical model.  The signal and simulated noise are summed, resulting in an image whose underlying signal and statistical properties agree with the original image.

Since statistical modeling of general multimedia data may be impractical, 'lossy' data compression methods may be applied to the noise function.  A successful lossy representation may be regarded as an alternate microstate of the perceived noise macrostate; it is effectively another sample drawn from the set of data macroscopically equivalent to the observed datum\cite{me1}.  As such, in the absence of an appropriate model, the noise function is compressed using lossy methods, normalized such that the resulting intensities do not exceed the maximum possible amplitude.  This representation will be decompressed and added to the decompressed signal to reconstruct the original datum.

This has several advantages.  First, the signal is relatively free of spurious artifacts, such as ringing, which interfere with the extraction of useful inferences from this signal.  Artifacts from lossy compression cannot exceed the amplitude of the noise floor, and higher levels of lossy compression may be used as a result of this fact.  Furthermore, lossy compression algorithms tend to compress high frequencies at a higher level than lower frequencies.  The eyes and ears tend to sense trends that exhibit change over broader regions of space or time, as opposed to high-frequency oscillations.  The compressibility of signal and noise leads to an information-theoretic reason for this phenomenon - the former naturally requires less of the nervous system's communication bandwidth than the latter.

The compression ratios afforded by such a scheme can be dramatic for noisy data.  As a trivial example, consider a string containing only random noise, such as the result of $n$ independently distributed Bernoulli trials having probability $p=0.5$, such as a coin flip.  Lossless entropic compression cannot effectively compress such a string.  Decomposing such a string into basis functions, such as the Fourier amplitudes or wavelets used in the JPEG algorithms, inevitably results in a mess of spurious artifacts with little resemblance to the original string.  The critical compression scheme described, however, easily succeeds in reproducing noise that is statistically indistinguishable from (though not identical to) the original string.  Furthermore, all that needs to be stored to sample from this distribution is the probability $p=0.5$ of the Bernoulli trial, which has complexity O(1).  The observer for which this scheme is optimal makes statistical inferences of amplitude in a manner similar to a physical measurement.  The observer records the statistics of the data, e.g. mean, variance, etc., rather than encoding particular data, which could introduce bias.

If the data is a waveform sampled at a frequency exceeding its effective Nyquist rate, such as an audio recording sampled at more than twice the frequency of a listener's ear, then its spectrum may be analyzed rather than its time series.  This will make the data smoother and non-negative, resulting in better compression for the leading bits.  In practical terms, this means that we may compress audio by compressing an one-dimensional image which is a graph of its spectrum, or the spectrum of some portion of the time series.  Hence, we will develop the method using images as a canonical example, with the understanding that audio may be compressed, for example, using 1-d images of Fourier transforms, and that video may be compressed using an array having the third dimension of time, or by embedding information into lower-dimensional arrays.

\subsection{Rotating the Color or Sample Space}

For many photographic and video applications, it is conventional to rotate a pixel's RGB color space to a color space, $YC_{b}C_{r}$, which more naturally reflects the eye's increased sensitivity to brightness as compared to variations in color.  This is normally done in such a way that takes into account the perceived variations in brightness between different phosphors, inks, or other media used to represent color data.

The $Y$ or luma channel is a black-and-white version of the original image which contains most of the useful information about the image, both in terms of human perception and measurements of numerical error.  The luma channel could be said to be the principal component (or factor) of an image with respect to perceptual models.  The blue and red chroma channels ($C_{b}$ and $C_{r}$, respectively) effectively blue-shift and/or red-shift white light of a particular brightness; they are signed values encoding what are typically slight color variations from the luma channel.  It is conventional for the luma channel to receive a greater share of bandwidth than the less important chroma channels, which are often downsampled or compressed at a lower bitrate.

As an alternative to consistently using a perceptual model optimized for the output of, for example, particular types of monitors or printers, one could use a similar approach to determine the principal components of color data as encoded rather than perceived.  Principal components analysis, also called factor analysis, determines linear combinations of inputs which have the most influence over the output.  In principal components analysis, $n$ samples of $m$-channel sample data are placed in the columns of an $m$-by-$n$ matrix A and the matrix product $AA^T$ is constructed to obtain an $m$-by-$m$ matrix.  The eigenvectors of $AA^T$ having the largest eigenvalues are the most influential linear combinations of data, the magnitude of these eigenvalues (sometimes called factor weights) reflects the importance of a particular combination.

The result of applying principal components analysis to photographic content leads to a customized color space whose principal component is a luma channel whose channel weights correspond to the eigenvector having the largest eigenvalue.  This channel is best communicated at a higher bitrate than the secondary and tertiary components, which are effectively chroma channels.  In the appendix, we will compare the results of critically compressing photographs in RGB format against compression using a critical luma channel with lossy chroma channels.  For most of these photographs, a critically compressed luma channel leads to more efficient representations than using only lossy wavelet transformations.

In general, perceived output may be optimized by analyzing the principal components of perceived rather than raw data.  In contrast, directly applying principal component analysis (or factor analysis) to the raw data leads to a universal coordinate system for sample space which has improved compressibility, albeit optimized for a particular instance of data rather than the perceived output of a particular medium.  In addition to improved compression, another advantage of this approach is that it applies to a wide variety of numerical data and this facilitates a general approach to lossy data compression.

\section{Critical Bit Depth}

The critical bit depth determines the level of compressible content of a signal.  We now determine expressions for the first and second order critical depths.  This will allow us to separate signal from noise for audio, image, and video data by determining a bit depth that effectively separates signal from noise.  If higher compression ratios are desired, a supercritical signal may be used, meaning that more bits may be truncated, at the cost of destroying compressible information and potentially impeding inference.  On the other hand, a signal retaining nearly all of its bits would necessarily be similar to the original.

For a string which encodes the outcome of a series of independent Bernoulli trials (coin flips, for instance) as zeros and ones, each bit constitutes the same amount of information - one bit is one sample.  For a string comprised of a series of numeric samples at a bit depth greater than one, this is not usually the case.  In the traditional representation of numerals, leading bits are generally more informative than trailing bits, so an effective lossy data compression scheme should encode leading bits at a higher rate.  From the viewpoint of compressibility, on the other hand, the smooth, redundant leading bits of a typical stochastic processes are more compressible than its trailing bits.  Since the leading bits of multi-bit samples are often more compressible and more significant than the trailing bits, they are candidates for exact preservation using lossless data compression.  Since the trailing bits are generally less important and also less compressible, lossy compression can greatly reduce their descriptive complexity without perceptible loss.

\subsection{Bit Depth of a 2-D channel}

Since images will be easy to illustrate in this medium, and provide a middle ground as compared to one or three dimensions for audio or video, respectively, we will treat the two-dimensional case first.  We will then generalize to data having any number of dimensions.  We will refer to the matrices (rank-2 tensors) as images, since this is a canonical and intuitive case, but these expressions apply generally to all two-dimensional arrays of binary numbers.

Let $X^d_{i,j}$ represent a tensor of rank three (a tensor of rank n is an n-dimensional array) representing one channel of a bitmap image.  Subscript indices $i$ and $j$ represent $x$ and $y$ coordinates in the image, and the superscript $d$ indexes the bits encoding the amplitude of pixel (i,j) in the channel, ordered from most significant to least significant.

Let the set $A^n$ contain all the images whose n leading bits agree with those of $X^d_{i,j}$:
\begin{equation}
A^n \equiv \{Y^m_{ij} : Y^m_{ij} = X^m_{ij}, m \leq n \leq d \}
\end{equation}

This set can be represented as the original image channel with bit depth truncated from $d$ to $n$.  The implied observer sees n significant bits of (learnable) signal and $d-n$ insignificant bits of (non-learnable) noise.  For reference, the algorithmic entropy of the truncated string is:
\begin{equation}
H(Z) = K(A^n) + \log |A^n| = K(A^n) + ijd - ijn
\end{equation}

The literal length of the reduced image is $ijd-ijn$, and most of this will be saved in a critical compression scheme, as noise can be coded at a high loss rate.  If $B^n$ is the lossy representation, the complexity of the critically compressed representation is:

\begin{equation}
K = K(A^n) + K(B^n) \leq K(A^n) + ijd - ijn
\end{equation}

We may now consider the Fisher information (and hence the minimal sufficient statistic) of $A^n$.  The Fisher information of a path from $n$ to $n+1$ is, replacing the continuous differentials with finite differences, and ignoring the expectation operator (which becomes equivalent to the identity operator):
\begin{equation}
I(n) = 2 \ln 2\ (K(A^{n+1})-K(A^n))^2, 0<n<d
\end{equation}

The first order bit depth $n_0$ parameterizing the minimal sufficient statistic $A^{n_0}$ is the argument n that maximizes the change in complexity, $K(A^{n+1})-K(A^n)$:
\begin{equation}
n_0 = \arg \max I(n), 0<n<d
\end{equation}

The first order bit depth of the image channel represented by $X^d_{i,j}$ is $n_0$.  That is, the first $n_0$ most significant bits in each amplitude encode the signal; the remaining bits are noise.  The noise floor of the image is $n_0$.

The second order depth, on the other hand, determines the point of diminishing returns beyond which further description has diminished utility.  It is the maximum of the expected Hessian of Hartley information, $J(t)_{ij} = E[\partial^2_{t_i} \ln{f(X;t)} \partial^2_{t_j} \ln{f(X;t)}]$, so it becomes:
\begin{equation}
n_c = \arg \max J(n), 0<n<d
\end{equation}

This minimizes $K(A^{n+1}) - 2K(A^{n}) + K(A^{n-1})$.  The signal having $n_c$ bits per sample has the high-value bits and the residual noise function contains the bits determined to have diminishing utility.

\subsection{Critical Depths of Multi-Channel Data}

When considering multiple channels at once, which allows data compression to utilize correlations between these channels, we simply consider a superset of $A^n$ that is the union of the $A^n$ for each channel $X_k$, $0 \leq k < k_{max}$.  If all the channels have the same bit depth, for instance, this superset becomes:
\begin{equation}
A^n \equiv \{\bigcup Y^m_{ij} : Y^m_{ij} = (X_k)^m_{ij}, m \leq n \leq d, 0 \leq k < k_{max}\}
\end{equation}

Its corresponding representation is the union of the truncated channels, traditionally, an image would have three of these.  Given this new definition, the calculation of first-order depth proceeds as before. Its Fisher information is still $I(n) = 2 \ln 2 (K(A^{n+1})-K(A^n))^2$, which takes its maximum at the minimal sufficient statistic, the first order depth maximizing $K(A^{n+1})-K(A^n)$.  The second order depth, as before, maximizes $K(A^{n+1}) - 2K(A^{n}) + K(A^{n-1})$.

It is also possible to take the union of channels having different bit depths.  The first-order critical parameters (bit depths) are best determined by the maximum (or a near-maximum) of the Fisher-Rao metric.  The second-order critical parameters are determined by the maximum (or a near-maximum) of the Hessian of Hartley information.

\subsection{Representations using Multiple Critical Depths}

If stochastic data is not ergodic, that is, if different regions of data have different statistical properties, these regions may have different critical bit depths.  Such data can be compressed by separating regions having different bit depths.

This phenomenon occurs frequently in photographs since brighter regions tend to be encoded using more significant bits, requiring fewer leading bits.  Brighter regions thus tend to have lower critical depth than darker regions whose signal is encoded using less significant bits.  Darker regions require a greater number of leading bits, but their leading zeros are highly compressible.

The simplest way to accomplish this separation, perhaps, is to divide the original data into rectangular blocks (see the example) and evaluate each block's critical depth separately.  This is suboptimal for a couple of reasons - one, regions of complex data having different bit depths are rarely perfect rectangles; two, normalization or other phenomena can lead to perceptible boundary effects at the junctions of blocks.

For this reason, we develop a means of masking less intense signals for subsequent encoding at a higher bit depth.  In this way, the notion of bit depth will be refined - by ignoring leading zeros, the critical bit depth of the data becomes the measure of an optimal number of significant figures (of the binary fraction $2^{-x} = 0.x$) for sampled amplitudes.

Ideally, we would like to encode the low-amplitude signals at a higher bit depth (given their higher compressibility) while we make a lossy approximation of the Fourier transform of the periodic noise.  If a statistical model is available for this approximation, we use this model for the lossy coding, otherwise, a lossy data compression algorithm is employed.

Given the original data, it is easy to distinguish low-amplitude signals from truncated noise - if the original amplitude is greater than the noise floor, a pixel falls into the latter category, otherwise, the former.  This allows us to create a binary mask function $M^d$ associated with bit depth d, it is 0 if the amplitude of the original data sample exceeds the noise floor, and 1 otherwise:
\begin{equation}
M^n_{ij} = 0\ if A^n_{ij} \geq 2^d;\\
M^n_{ij} = 1\ if A^n_{ij} < 2^d.\\
\end{equation}

This mask acts like a diagonal matrix that left-multiplies a column vector representation of the image $A^d_{ij}$.  The resulting signal $M^d_{ij}A^d_{ij}$ preserves regions of non-truncated low-intensity signal while zeroing all all other amplitudes.  Its complement, NOT $M^d_{ij} = \bar{M}^d_{ij}$ acts on the noise function to preserve regions of periodic truncated noise while zeroing the low-intensity signal.  It is also helpful to consider the literal description length of the samples contained in this region, its bit depth times the number of ones appearing in $M^n$: $(d-n)\ trace(M^n)$, as well as the complexity of the entire signal, $K(M^d_{ij}A^d_{ij})$, which includes the shape of the region, since this is additional information that needs to be represented.

We may now describe an algorithm that calculates critical depths while separating regions of low-intensity signal.  This procedure truncates some number (the bit depth) of trailing digits from the original signal, separates truncated regions having only leading zeros, and calculates the complexity of the reduced and masked signal plus the complexity (calculated via recursion) of the excised low-intensity signal at its critical depth.  Once this is done, it proceeds to the next bit depth, provided that the maximum depth has not yet been reached. %optimizations

Starting with shallow representations having only the most significant bits (n=0 or 1, typically), we truncate the data to depth n, resulting in the truncated representation of the signal $A^n$, as well as its truncated residual (previously 'noise') function, which we will call $B^n$.  At this point, the initial truncated signal $A^n$ is compressed using lossless methods, while the mask and its complement are applied to $B^n$.  This results in a residual signal, $S^n = M^n B^n = M^n A^d_{ij}$ (for these pixels, $B^n$ agrees with the original data $A^d$) as well as a complementary residual periodic noise function $N^n = \bar{M}^n B^n$.  Since it contains only residual noise, taken modulo some power of two, the noise $N^n$ is compressed using lossy methods that are typically based on Fourier analysis.  The residual signal $S^n$ becomes input for the next iteration.

The procedure iterates using a new $A^{n+1}$ that is a truncation of the masked signal $S^n$, as opposed to the original image.  Let the notation $T^n$ represent an operator that truncates amplitudes to bit depth n.  In this notation, $A^{n+1} = T^{n+1} S^n$, and its residual function is $B^{n+1} = A^{n}-A^{n+1}$.  A new mask $M^{n+1}$ is determined from $A^{n+1}$.  Using the new mask, we produce a new residual signal, $S^{n+1} = M^{n+1} B^{n+1}$, and a new residual noise, $N^{n+1} = \bar{M}^{n+1} B^{n+1}$.  $A^{n+1}$ is compressed and stored using lossless methods, while $N^{n+1}$ is compressed and stored using lossy methods, and the procedure iterates to the next value of n, using $S^{n+1}$ as the new signal provided that additional bits exist.  If the maximum bit depth has been reached, there can be no further iteration, so $B^n$ is stored using lossy methods.

Though the separation of signal and noise is now iterative, the criterion for critical depth has not changed, only the $A^n$ that appears in their definition. If $K(A^n)-K(A^{n+1})$ is nearly maximal - its largest possible value is the literal length, $trace(M^n)$ - the first-order depth has been reached; if $K(A^{n-1})- 2 K(A^{n}) + K(A^{n+1})$ is nearly maximal, the second-order critical depth has been reached.  Once the desired depth is reached, the iteration may break, discarding $S^{n+1}$ and $N^{n+1}$ and storing $B^n$ using lossy methods.

If higher compression ratios are desired, more bits can be truncated and modeled statistically or with lossy methods.  However, the signal introduced to the noise function in such a manner might not fit simple statistical models, and the loss of compressible signal tends to interfere with inference.

\section{Critical Scale}

Since it relates an image's most significant bits to important theoretical notions such as Kolmogorov minimal sufficient statistic and algorithmic entropy, critical bit depth is the canonical example of critical data representation.  One could reorder the image data to define a truncated critical scale, simply sampling every nth point, however, this discards significant bits, which tends to introduce aliasing artifacts dependent on the sampling frequency and the frequency of the underlying signal.  These considerations are the topic of the celebrated Nyquist-Shannon sampling theorem - essentially, the sampling frequency must be at least twice the highest frequency in the signal.  This is known as the Nyquist rate, as it was stated by Nyquist in 1928\cite{ny28} before finally being proved by Shannon in 1949\cite{sh49}.

As a result of the Nyquist-Shannon theorem, a downsampling operation should incorporate a low-pass filter to remove elements of the signal that would exceed the new Nyquist rate.  This should occur prior to the sampling operation in order to satisfy the sampling theorem.  To complicate matters further, idealized filters can't be attained in practice, and a real filter will lose some amount of energy due to the leakage of high frequencies.  A low-pass filter based on a discrete Fourier transform will exhibit more high-frequency leakage than one based on, for example, polyphase filters.  Since sampling is not the topic of this paper, we will simply refer to an idealized sampling operator $B$ that applies the appropriate low-pass filters to resample data in one or more dimensions.

The ability to perform complexity-based inference on a common scale is important since it allows the identification of similar objects, for instance, at different levels of magnification.  Critical scale is useful as it provides another degree of freedom along which critical points may be optimized, analogous to phase transitions in matter that depend on both temperature and pressure.  Occasionally, the two objectives may be optimized simultaneously at a 'triple point' that is critical for both bit depth and scale.

We now define the critical scale, which we define in terms of operators which resample the image instead of truncating it.  For some data, a minimal sufficient statistic for scale cannot be reliably selected or interpreted, and hence a critical scale can't be determined.

Consider the critical scale of an image at a particular bit depth $d$, which may or may not be the original bit depth of the image.  Let the linear operator $B_{r,s}$ represent a resampling operation, as described above, applied to an image with a spatial period of r in the x dimension and a period of s in the y dimension.  It's action on $X^d_{i,j}$ is a $i/r$ by $j/s$ matrix of resampled amplitudes.

This operator gives us two possible approaches to scaling.  On one hand, given divisibility of the appropriate dimensions, we may vary $r$ and $s$ to resample the image linearly.  On the other hand, we may also apply the operator repeatedly to resample the image geometrically, given divisibility of the dimensions by powers or r and s.  The former may identify the scale of vertical and horizontal components separately, and the latter identifies the overall scale of the image.  We will consider first overall scale, using the iterated operator, and then the horizontal and vertical components.

$B_{r,s}^n$, then, is the result of applying this operator n times.  Let the set used by the structure function, $A^n$, contain all the images whose m leading bits agree with those of $B_{r,s}^n X^d_{i,j}$:
\begin{equation}
A^n \equiv \{Y^m_{ij} : Y^m_{ij} = B_{r,s}^n X^m_{ij}, m \leq d \}
\end{equation}

Note that in this case, unlike the critical bit depth, $A^n$ is an averaging which is not necessarily a prefix of the original image.  The original image has $ijd$ bits.  The reduced image has $\frac{dij}{(rs)^n}$ bits.  We may now write the first-order critical scale $n_0$ parameterizing the minimal sufficient statistic $A^{n_0}$, an expression unchanged from the previous case:
\begin{equation}
n_1 = \arg \max I(n) = \arg \max 2 \ln 2 (K(A^{n+1})-K(A^n))^2, 0<n<d
\end{equation}

If higher compression ratios are needed, additional signal may be discarded, as described previously.  The expression for the second-order critical scale is also unchanged:
\begin{equation}
n_2 = \arg \max J(n) = \arg \max K(A^{n+1}) - 2K(A^{n}) + K(A^{n-1}), 0<n<d
\end{equation}

As mentioned previously, the repeated application of averaging operators is not always appropriate or possible.  We will consider linear scaling parameterized along the horizontal axis, with the understanding that the same operations may be applied to the vertical axis, or to any other index.  As such, we equate the parameter $n$ with the horizontal factor $r$.  The set $A^n$ then contain all the images whose m leading bits agree with those of $B_{n,s} X^d_{i,j}$:
\begin{equation}
A^n \equiv \{Y^m_{ij} : Y^m_{ij} = B_{n,s} X^m_{ij}, m \leq d \}
\end{equation}

Note that this set may not be defined for all n.  Given this set, the expressions for the maximum Fisher information (the minimal sufficient statistic) and the maximum of the Hessian of Hartley information (the critical point) do not change.

If a signal is to compressed at some scale other than its original scale, then it will need to be resampled to its original scale before being added to its decompressed (lossy) noise function.  Note than in this case, the resampled signal is not generally lossless.  This smoothing of data may be acceptable, however, since it enables analysis at a common scale.

\section{Multidimensional, Multi-channel Data}

We now consider critical bit depths and scales of multidimensional data.  Instead of a two dimensional array containing the amplitudes of an image, we consider an array with an arbitrary number of dimensions.  As noted earlier, monophonic audio may be represented as a one dimensional array of scalar amplitudes, and video data may be represented as a three dimensional array which also has three channels.  This generalizes the results of the previous two sections, which used the case of a two-dimensional image for illustrative purposes.

Let $X^{d_0 ... d_{v}}_{a_0 ... a_{m-1}}$ represent a tensor of rank $m+v$.  Its subscripts index coordinates in an $m$-index array whose values are $v$-dimensional vectors.  Each vector component is a $d_i$-bit number.  The superscripts index the bits in these numbers, ordered from most significant to least significant.  We will first determine its critical bit depths and then its critical scales.

Let the set $A^{n_0 ... n_{v}}$ contain all possible tensors $Y^d_{a_0 ... a_{m-1}}$ whose whose $n_i$ leading bits agree with those of channel $i$ in the original tensor $X^{d_0 ... d_{v}}_{a_0 ... a_{m-1}}$:
\begin{equation}
A^{n_0 ... n_{v}} \equiv \{Y^{b_0 ... b_{m-1}}_{a_0 ... a_{m-1}} : Y^{b_0 ... b_{m-1}}_{a_0 ... a_{m-1}} = X^{b_0 ... b_{m-1}}_{a_0 ... a_{m-1}}, b_i \leq n_i \leq d_i \}
\end{equation}

Since there are multiple parameters involved in determining the first-order bit depth, a general tensor requires the use of the full Fisher-Rao metric, rather than a Fisher information:
\begin{equation}
I(n_0 .. n_{v})_{ij} \sim ( K(A^{n_0 .. n_{i+1} .. n_{v}})-K(A^{n_0 .. n_{i} .. n_{v}}) )
( K(A^{n_0 .. n_{j+1} .. n_{v}})-K(A^{n_0 .. n_{j} .. n_{v}}) )
\end{equation}

Where the expectation value in the definition of I(t) becomes the identity, as before.  For any particular parameter $n_k$, $I(t)$ takes on a maximum with respect to some value that parameter.  This value is the critical depth of channel $k$.  However, this does not necessarily indicate that the set of critical depths globally maximizes I(t).  The global maximum occurs at some vector of parameter values $\vec{n} = (n_0, ..., n_{v})$:
\begin{equation}
\vec{n_1} = \arg \max I(n_0 ... n_{v})_{ij}, 0<n_i<d
\end{equation}

If channels having different depths are inconvenient, a single depth may be selected as before, using the one-parameter Fisher information of $P(X)=2^{-K(X)}$:
\begin{equation}
n_1 = \arg \max I(n)_{ij} = \arg \max 2 \ln 2 (K(A^{n+1})-K(A^n))^2, 0<n<d
\end{equation}

The second-order critical depth proceeds in a similar manner.  The expected Hessian of Hartley information $J(\vec{n})_{ij}$ becomes $2 \ln 2$ times
\begin{equation}
( K(A^{...n_{i+1}...})-2K(A^{...n_{i}...}) + K(A^{...n_{i-1}...}) )
( K(A^{...n_{j+1}...})-2K(A^{...n_{j}...})+K(A^{...n_{j-1}...}) )
\end{equation}

Again, the maximum occurs at a vector of parameter values $\vec{n_c} = (n_0, ..., n_{v})$:
\begin{equation}
\vec{n_2} = \arg \max J(n_0 ... n_{v})_{ij}, 0<n_i<d
\end{equation}

We will now consider the critical scale of $X^d_{a_0 ... a_{m-1}}$.  Let the linear operator $B_{i_0...i_{m-1}}$ represent an idealized resampling of the tensor by a factor of $i_k$ along each dimension $k$.  This operator applies a low-pass filter to eliminate frequency components that would exceed twice the new sampling frequency (the Nyquist rate $\frac{2}{i_k}$) prior to sampling with frequency $\frac{1}{i_k}$.

Let the set used by the structure function, $A^{n_0 ... n_{v}}$, contain all the tensors which are preimages of $B_{i_0...i_{m-1}} X^d_{i,j}$, the rescaled tensor:
\begin{equation}
A^{n_0 .. n_{v}} \equiv \{Y^{b_0 ... b_{m-1}}_{a_0 ... a_{m-1}} : Y^{b_0 ... b_{m-1}}_{a_0 ... a_{m-1}} = B_{i_0...i_{m-1}}^n X^{b_0 ... b_{m-1}}_{a_0 ... a_{m-1}}, m \leq n \leq d \}
\end{equation}

Given this new definition of A, the definition of first and second order critical depth do not change.  The first-order critical scales maximize (or approximately maximize) the Fisher information:
\begin{equation}
\vec{n} = \arg \max I(n_0 ... n_{v})_{ij}
\end{equation}

Likewise, the second-order critical scales maximize the expected Hessian of Hartley information:
\begin{equation}
\vec{n} = \arg \max J(n_0 ... n_{v})_{ij}
\end{equation}

Alternately, a single scale parameter could be chosen in each case:
\begin{equation}
n_1 = \arg \max I(n ... n)_{ij} = \arg \max K(A^{n+1})-K(A^n), 0<n<d
\end{equation}
\begin{equation}
n_2 = \arg \max J(n ... n)_{ij} = \arg \max K(A^{n+1})-2K(A^n)+K(A^{n+1}), 0<n<d
\end{equation}

Hence, we see that the definition of first and second order critical scale of a general tensor is identical to their definition for rank two images.

The above relations are idealized, assuming that K can be evaluated using perfect data compression.  Since this is not generally the case in practice, as discussed previously, the maxima of I and J may be discounted by some tolerance factor to produce the threshold of a set of effectively maximal parameter values.  The maximum or minimum values within this set may be chosen as critical parameters.

In addition to multimedia data such as audio (m=1), images (m=2), and video (m=3), these relations enable critical compression and decompression, pattern recognition, and forecasting for many types of data.

\section{Modeling and Sampling Noise}

We now have an approach to separate significant signals from noise.  Encoding the resulting signal follows traditional information theory and lossless compression algorithms.  Encoding the noise is a separate problem of statistical inference that could assume several variants depending on the type of data involved as well as its observer.  Regardless of the details of the implementation, the program is as follows: a statistical model is fit to the noise, its parameters are compressed and stored, and upon decompression, a sample is taken from the model.

A lossless data compression scheme must encode both a signal and its associated noise.  However, noise is presumed ergodic and hence no more likely or representative than any other noise sampled from the same probability distribution.  Hence, a 'lossy' perceptual code is free to encode a statistical model (presuming that one exists) for the noisy bits, as their particular values don't matter.  This dramatically reduces the complexity of storing incompressible noise; its effective compression ratio may approach 100\% by encoding relatively simple statistical models.  The most significant bits of the signal are compressed using a lossless entropic code; when the image is decompressed, samples are taken from the model distribution to produce an equivalent instance of noise; this sampled noise is then added to the signal to produce an equivalent image.

As noted in the introduction, maximum likelihood estimates correspond to sufficient statistics\cite{FI21}.  Maximum likelihood estimation (MLE) has been one of the most celebrated approaches to statistical inference in recent years.  There are a wide variety of probability distribution functions and stochastic processes to fit to data, and many specialized algorithms have been developed to optimize the likelihood.  A full discussion of MLE is beyond the scope of this paper, but the basic idea is quite simple.  Having computed a minimal sufficient statistic, we wish to fit a statistical model having some finite number of parameters (such as moments of the distribution) to the noisy data.  The parameter values leading to the most probable data are selected to produce the most likely noise model.  Our data compression scheme stores these parameter values, compressed losslessly if their length is significant, in order to sample from the statistical model when decompression occurs.

Since different regions of data may have different statistical properties, rather than fitting a complex model to the entire noisy string, it may be advantageous to fit simpler models to localized regions of data.  The description length of optimal parameters tends to be proportional to the number of parameters stored.  If a model contains too many parameters, the size of their representation can approach the size of the literal noise function, reducing the advantage of critical compression.

When the image is decompressed, a sample is drawn from the statistical model stored earlier.  The problem of sampling has also received considerable attention in recent years, due to its importance to so-called Monte Carlo methods.  The original Monte Carlo problem, first solved by Metropolis and Hastings, dealt with the estimation of numerical integrals via sampling.  Since then, Monte Carlo has become somewhat of a colloquial term, frequently referring to any algorithm that samples from a distribution.  Due to the importance of obtaining a random sample in such algorithms, Monte Carlo sampling has become a relatively developed field.  Box-Muller is a simple option for sampling from the uniform distibution\cite{Knuth2}, which may be transformed into any other distribution with a known cumulative distribution function.  The Ziggurat algorithm\cite{MT00} is a popular sampling algorithm for arbitrary distributions.  This algorithm provides reasonably high-performance sampling since the sequence of samples produced is known to repeat only after a very large number of iterations.

Psychological modeling should be incorporated into the statistical models of the noise component rather than the signal, since this is where the 'loss' occurs in the compression algorithm.  Since noise can't be learned efficiently, different instances of noise from the same ergodic source will typically appear indistinguishable to a macroscopic observer who makes statistical inferences.  Furthermore, certain distinct ergodic sources will appear indistinguishable.  A good psychological model will contain parameters relevant to the perception of the observer and allow irrelevant quantities to vary freely.  It may be advantageous to transform the noisy data to an orthogonal basis, for example, Fourier amplitudes, and fit parameters to a model using this basis.  The particular model used will depend on the type of data being observed and, ultimately, the nature of the observer.  For example, a psychoacoustic model might describe only noise having certain frequency characteristics.   The lossy nature of the noise function also provides a medium for other applications, such as watermarking.

Maximum likelihood estimation applies only when an analytic model of noise is available.  In the absence of a model, the noise function may be compressed using lossy methods, as described previously, and added to the decompressed signal to reconstruct the original datum.

The most obvious obstacle to this procedure is the fact that lossy data compression algorithms do not necessarily respect the intensity levels present in an image.  For example, high levels of JPEG compression produces blocking artifacts resembling the basis functions of its underlying discrete cosine transformation.  Fitting these functions to data may result in spurious minima and maxima, often at the edges or corners of blocks, which frequently exceeds the maximum intensity of the original noise function by nearly a factor of two.  With the wavelet transforms used in JPEG2000\cite{AT05}, the spurious artifacts are greatly diminished compared to the discrete cosine transforms of the original JPEG standard, but still present, so a direct summation will potentially exceed the ceiling value allowed by the image's bit depth.

In order to use the lossy representation, the spurious extrema must be avoided.  It is not appropriate to simply truncate the lossy representation at the noise floor, as this leads to 'clipping' effects - false regularities in the noise function that take the form of artificial plateaus.  Directly normalizing the noise function does not perform well, either, as globally scaling intensities below the noise floor leads to an overall dimming of the noise function relative to the reconstructed signal.  A better solution is to upsample the noise function, normalizing it to the maximum amplitude, and storing the maximum and minimum intensities of the uncompressed noise.  Upon decompression, the noise function image is 'de-normalized' (downsampled) to its original maximum and minimum intensities before being summed with the signal.  This process preserves the relative intensities between the signal and noise components.

Therefore, when constructing codes with both lossless and lossy components, a normalized (upsampled) noise function is compressed with lossy methods, encoded along with its original bit depth.  Upon decompression, the noise function is decompressed and re-normalized (downsampled) back to its original bit depth before being added to the decompressed signal.

\section{Artificial Intelligence}

Last, but not least, it is worth noting that separating a signal from noise generally improves machine learning and pattern recognition.  This is especially true in compression-based inference.  Complexity theory provides a unified framework for inference problems in artificial intelligence\cite{LV97}, that is, data compression and machine learning are essentially the same problem\cite{CO91} of knowledge representation.

These sorts of considerations have been fundamental to information theory since its inception.  In recent years, compression-based inference experienced a revival following Rissanen's 1986 formalization\cite{LV97} of minimal description length (MDL) inference.  Though true Kolmogorov complexities (or algorithmic prefix complexities) can't be calculated in any provable manner, the length of a string after entropic compression has been demonstrated as a proxy sufficient for statistical inference.  Today, data compression is the central component of many working data mining systems.  Though it has historically been used to increase effective network bandwidth, data compression hardware has improved in recent years to meet the high-throughput needs of data mining.  Hardware solutions capable of 1 gbps or more of Lempel-Ziv compression throughput can be implemented using today's FPGA architectures.  In spite of its utility in text analysis, compression-based inference remains largely restricted to highly compressible data such as text.  The noise intrinsic to stochastic real-world sources is, by definition, incompressible, and this tends to confound compression-based inference.

Essentially, incompressible data is non-learnable data\cite{LV97}; removing this data can improve inference beyond the information-theoretic limits associated with the original string.  By isolating the compressible signal from its associated noise, we have removed the obstacle to inference using the complexity of stochastic data.  Given improved compressibility, the standard methods of complexity-based inference and minimum description length may be applied to greater effect.

By comparing the compressibility of signal components to the compressibility of their concatenations, we may identify commonalities between signals.  A full treatment of complexity-based inference is beyond the scope of this paper (the reader is referred to the literature, particularly the book by Li and Vit\'{a}nyi\cite{LV97}) but we reproduce for completeness some useful definitions.  Given signal components A and B, and their concatenation AB, we may define the conditional prefix complexity $K(A|B) = K(AB)-K(B)$. This quantity is analogous to the Kullback-Leibler divergence or relative entropy, and this allows us to measure a (non-symmetric) distance between two strings.

Closely related to the problem of inference is the problem of induction or forecasting addressed by Solomonoff\cite{LV97}.  Briefly, if the complexity is K(x), then the universal prior probability $M(x)$ is typically dominated by the shortest program, implying $M(x) \approx 2^{-K(x)}$.  If x is, for example, a time series, then the relative frequency of two subsequent values may be expressed as a ratio of their universal probabilities.  For example, the relative probability that the next bit is a 1 rather than 0 is $P = \frac{M(x1)}{M(x0)} \approx 2^{K(x0)-K(x1)}$.  Clearly, the evaluation of universal probabilities is crucially dependent on compressibility.  As such, separating signal from noise in the manner described also improves the forecasting of stochastic time series.

Sometimes, one wishes to consider not only the complexity of transforming A to B, but also the difficulty of transforming B to A.  In the example presented, an image search algorithm, this is not the case - the asymmetric distances produce better and more intuitive results.  However, if one wishes to consider the symmetrized distances, the most obvious might be symmetrized sum of conditional complexities $K(A|B)$ and $K(B|A)$, $D(A,B) = K(AB)+K(BA)-K(A)-K(B)$.  This averaging loses useful information, though, and many authors\cite{LV97} suggest using the 'max-distance' or 'picture distance' $D(A,B) = \max\{K(A|B),K(B|A)\} = \max\{K(AB)-K(B),K(BA)-K(A)\}$.  When implementing information measures, the desire for the mathematical convenience of symmetric distance functions should be carefully checked against the nature of the application.  For example, scrambling an egg requires significantly fewer interactions than unscrambling and subsequently reassembling that egg, and a reasonable complexity measure should reflect this.  For these considerations, as well as its many convenient mathematical properties\cite{LV97} the conditional prefix complexity is often the best measure of the similarity or difference of compressible signals.  Empirical evidence suggests that conditional prefix complexity outperforms either the symmetrized mutual information or the max-distance for machine vision.  The reverse transformation should not usually be considered since this tends to overweight low-complexity signals.

We will demonstrate an example sliding-window algorithm which calculates conditional prefix complexities $K(A|B)$ between a search texture, A, and elements B from a search space.  The measure $K(A|B)=K(AB)-K(B)$ is closely related to the universal log-probability that A will occur in a sequence given that B has already been observed.  Its minimum is the most similar search element.  It is invariant with respect to the size of elements in the search space and does not need to be normalized.  Estimating the complexities of signal components from the search space and the complexities of their concatenations with signal components from the texture, we arrive at an estimate of $K(A|B)$ which may be effectively utilized in a wide variety of artificial intelligence applications.

The first-order critical point represents the level at which all the useful information is present in a signal.  At this point, the transition from order to chaos is essentially complete.  For the purposes of artificial intelligence, we want enough redundancy to facilitate compression-based inference, but alto to retain enough specific detail to differentiate between objects.  The second-order critical point targets the middle of the phase transition between order and chaos, where the transition is proceeding most rapidly and both of these objectives can be met.  In the examples section, we will see that visual inference at the second-order critical depth outperforms inference at other depths.  For this reason, second-order critically compressed representations perform well in artificial intelligence applications.

\section{Examples}

Having described the method, we now apply it to some illustrative examples.  We will start with trivial models and household data compression algorithms, proceeding to more sophisticated implementations that demonstrate the method's power and utility.  These examples should be regarded as illustrations of a few of the many possible ways to implement and utilize two-part codes, rather than specific limitations of this method.

One caveat to the application of complexity estimates is the fact that real-world data compression produces finite sizes for strings which contain no data, as a result of file headers, etc.  We will treat these headers as part of the complexity of compressed representations when the objective of the compression is data compression, as they are needed to reconstruct the data.  When calculating derivatives of complexity to determine critical points, this complexity constant does not affect the result, as the derivative of a constant is zero.  It does not affect the conditional prefix complexity, either, canceled by taking a difference of complexities.

For certain applications in artificial intelligence, however, small fluctuations in complexity estimates can lead to large difference in estimates of probabilities.  When the quality of complexity estimates are important, as is the case for inference problems, we will first compress empty data to determine the additive constant associated with the compression overhead, and this complexity constant is subtracted from each estimate of K(X).  Formally, the data compression algorithm is regarded as a computer and its overhead is absorbed into this computer's Turing equivalence constant.  This results in more useful estimates of complexity which improves our ability to resolve low-complexity objects under certain complexity measures.

First-order critical data compression represents a level of detail at which a signal is essentially indistinguishable from an original when viewed at a macroscopic scale.  We will show that second-order critical data compression produces representations which are typically slightly more lossy but significantly more compact.  As mentioned earlier, any depth could potentially be used, splitting a two-part code subject to the constraints of the intended application.

\subsection{Examples of Image Compression}

Since images may be displayed in a paper more readily than video or audio, we first consider the second-order critical depth of a simple geometric image with superimposed noise.  This is a 256x256 pixel grayscale image whose pixels have a bit depth of 8.  The signal consists of a 128x128 pixel square having intensity 15 which is centered on a background of intensity 239.  To this signal we add, pixel by pixel, a noise function whose intensity is one of 32 values uniformly sampled between -15 and +16.  Starting from this image, we take the n most significant bits of each pixel's amplitude to produce the images $A^n$, where n runs from 0 to the bit depth, 8.  These images are visible in figure 1, and the noise functions that have been truncated from these images are showcased in figure 2.

To estimate $K(A^n)$, which is needed to evaluate critical depth, we will compress the signal $A^n$ using the fast and popular gzip compression algorithm and compress its residual noise function into the ubiquitous JPEG format.  We will then progress to more accurate estimates using slower but more powerful compression algorithms, namely, PAQ8 and JPEG2000.  The results are tabulated below, with n on the left and the size of gzip's representation, the estimate of $K(A^n)$, in the right column.
\begin{center}
\begin{tabular}{cc}
  0 & 0 \\
  1 & 1584\\
  2 & 2176 \\
  3 & 13760 \\
  4 & 129200 \\
  5 & 282368 \\
  6 & 441888 \\
  7 & 659296 \\
  8 & 735520 \\
\end{tabular}
\end{center}

The most significant bits enjoy higher compression ratios than the noisy bits.  In addition to the relative smoothness of many interesting data, there are fundamental reasons (such as Benford's law and the law of large numbers) why leading digits should be more compressible.  To estimate the Kolmogorov sufficient statistic, we would to compress the original image in the order of the bits' significance until compressing an additional bit of depth increases the size of the representation just as much as storing that additional bit of depth without compression.  If the data compression algorithm doesn't see the image as being random (which is not the case here) we would store bits until adding an additional bit increases the representation by a maximal or nearly maximal amount.

A single image channel is second-order critical at depth n when $K(A^{n+1})-2K(A^n)+K(A^{n+1})$ is within some tolerance factor of its maximal value.  We select a 25 percent tolerance factor, chosen somewhat arbitrarily based on the typical magnitude of complexity fluctuations.  The depth n among this maximal set which has the minimum complexity $K(A^{n})$ is selected as the critical point.  For the given data, we see that no other second difference is within $25\%$ of the maximum, $n=3$.  The first-order critical point, on the other hand, doesn't occur until the phase transition is estimated to be complete at $n=7$.

\begin{center}
  \includegraphics[width=1.5in]{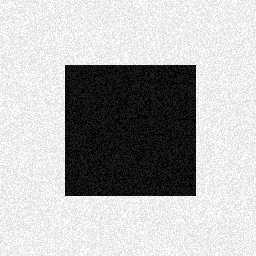}
  \includegraphics[width=1.5in]{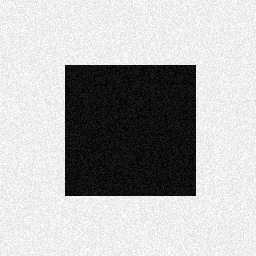}
  \includegraphics[width=1.5in]{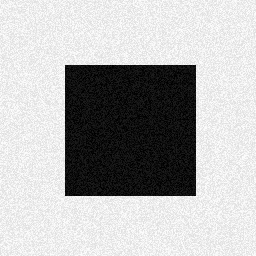}\\
  \smallskip
  \includegraphics[width=1.5in]{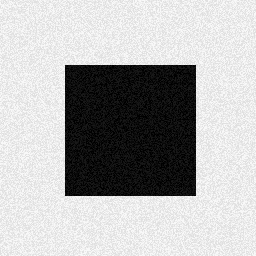}
  \includegraphics[width=1.5in]{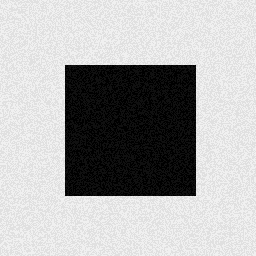}
  \includegraphics[width=1.5in]{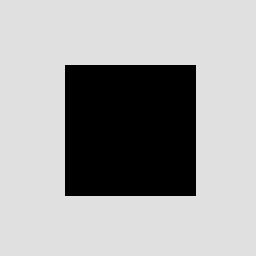}\\
   \smallskip
  \includegraphics[width=1.5in]{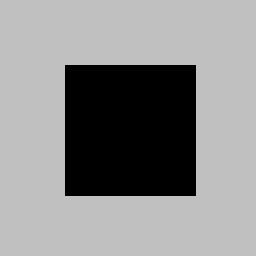}
  \includegraphics[width=1.5in]{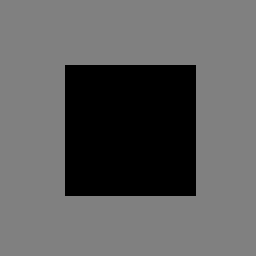}
  \includegraphics[width=1.5in]{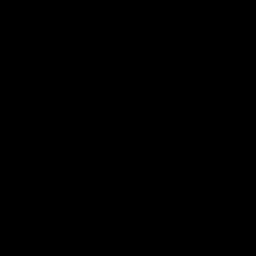}
\end{center}

Figure 1: Nine images of a dark gray box on a light gray background representing the signal at bit depths of 0 through 8.  The image at the upper left has all 8 bits, and the solid black image at the lower right represents the absence of any signal.  Note that depth 3, at middle right, is undergoing a phase transition between the previous five noisy images and the next three smooth images.  This is the second-order critical depth, the point of diminishing representational utility.

\bigskip

Having compressed the critical signal data, we may now fit a statistical model to the residual noise function, which may be seen in figure 2.  We will assume noise whose intensity function is discrete and uniformly distributed, as should be the case.  We use maximum likelihood estimation to determine the parameters, which are simply its minimum and maximum values.

\begin{center}
  \includegraphics[width=1.5in]{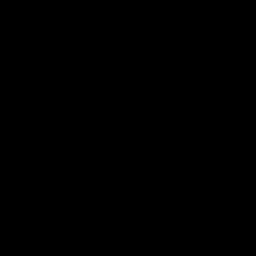}
  \includegraphics[width=1.5in]{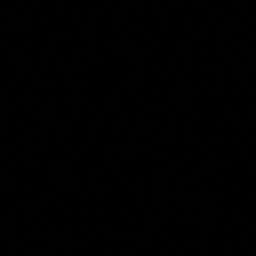}
  \includegraphics[width=1.5in]{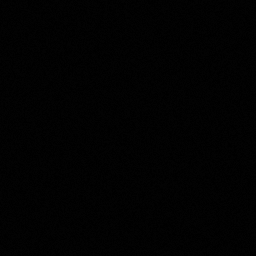}\\
  \smallskip
  \includegraphics[width=1.5in]{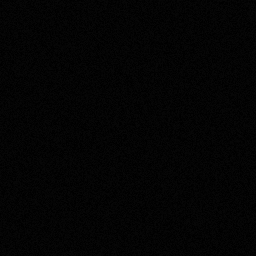}
  \includegraphics[width=1.5in]{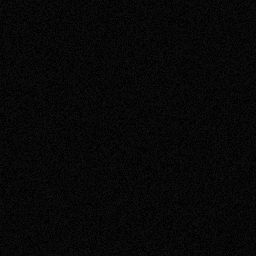}
  \includegraphics[width=1.5in]{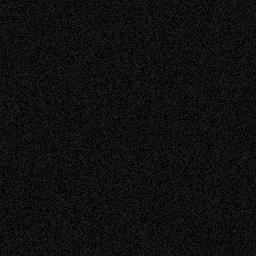}\\
  \smallskip
  \includegraphics[width=1.5in]{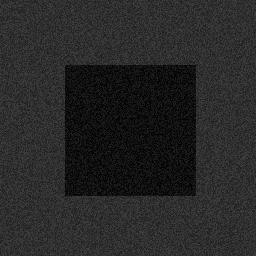}
  \includegraphics[width=1.5in]{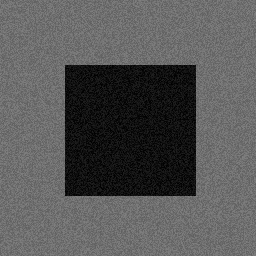}
  \includegraphics[width=1.5in]{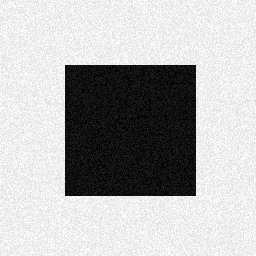}
\end{center}

Figure 2: Nine images representing the noise extracted from the signals illustrated in figure 1.  The first-order critical depth is depth 3, so the full noise function may be viewed at middle right.  It is clear that beyond this second-order critical depth, the signal of the image creeps back into the noise function in the lower row.

\bigskip

If the maximum and minimum values of the uniform discrete distribution are $a$ and $b$, the likelihood of intensity $i$ in any particular pixel is $\frac{1}{b-a}$ if $a \leq i \leq b$ and zero otherwise.  We assume pixels are independent, so the likelihood of any particular noise pattern is simply the product of the likelihoods of each pixel's noise.

To use maximum likelihood estimation, we vary $a$ and $b$ in order to determine which of their values leads to the maximum likelihood.  In practice, this can lead to very small numbers that suffer from floating-point errors, so it is customary to maximize the logarithm of likelihood, which is equivalent since the logarithm function is convex.  The log-likelihood is simply $a-b$ if $a \leq i \leq b$ and infinity otherwise.  This is summed over all 65536 pixels in the noise function and maximized.  The first-order critical depth is 3, so the bit depth of the noise function is 5, meaning that the noise can potentially take on one of $2^5=32$ values.  For brevity of presentation, we optimize by brute force, calculating noise function likelihoods for all possible a and b such that $-16 \leq b < a \leq 16$.  Executing this procedure, we see it recovers the correct set of parameters: $a=0$, $b=31$.  Note that this distribution has the same variance but a different mean from the noise generated for the test image.  A more appropriate optimization algorithm should have faster performance but the same result.

We have succeeded in applying second-order critical compression to the image.  We store $13760$ bits of compressed signal, corresponding to the second-order critical depth of 3, and 8 bits for each of the parameters $a$ and $b$ calculated above.  The compressed representation has $13776$ bits, whereas the original image has $524288$ bits and can't be losslessly compressed to such level without effectively cracking the pseudorandom number generator that was used to produce it.  Second-order critical compression attains a compression ratio of $38:1$ or $97.4\%$, and, as the reader may verify, the result is very difficult for the human eye to distinguish from the original image (try looking at the corner pixels) without a magnifying device.  For this reason, these images are referred to as being macroscopically equivalent; they are both microstates of the same macrostate.
\pagebreak
\begin{center}
  \includegraphics[width=2.25in]{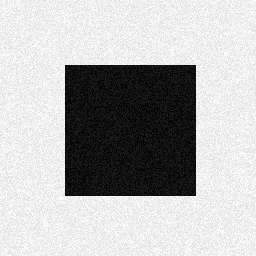}
  \includegraphics[width=2.25in]{signal8bit}\\
  \smallskip
  \includegraphics[width=2.25in]{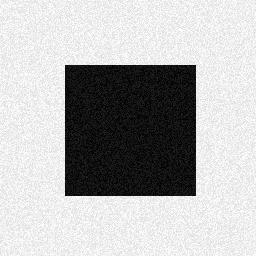}
\end{center}

Figure 3:  The second-order critically compressed image (13776 bits), at top left, adjacent to the uncompressed original image (524288 bits), at top right, for comparison.  The two images are practically indistinguishable to the naked eye.  Below, a JPEG compressed image (104704 bits) is displayed for comparison purposes.  In addition to being almost eight times the size of the critically compressed image, correlations in the form of blurring are visible in the JPEG-compressed noise, resulting in a less crisp textural appearance and biased statistical properties.

\bigskip

The previous example, a simple geometric signal in the presence of noise, was constructed to showcase the advantage of critical data compression.  We now perform a similar analysis using a color photograph.  We partition each of the image's red, green, and blue channels into their most and least significant bits.  The resulting channels may be viewed below, three per image.  This time we omit the no-signal (n=0 signal) image for brevity, as the absence of a signal simply appears black.
\begin{center}
  \includegraphics[width=4.7in]{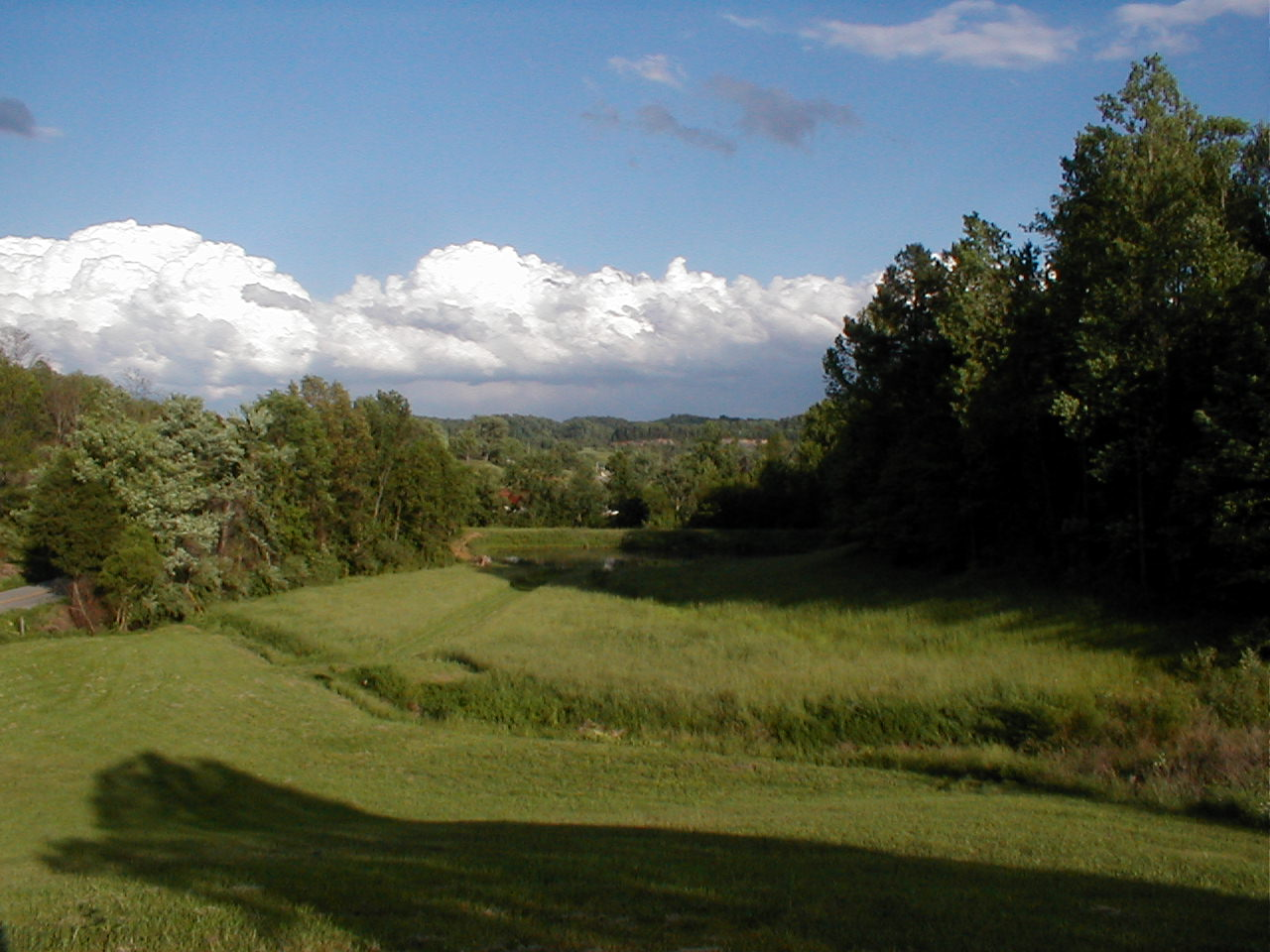}\\
\bigskip
  \includegraphics[width=4.7in]{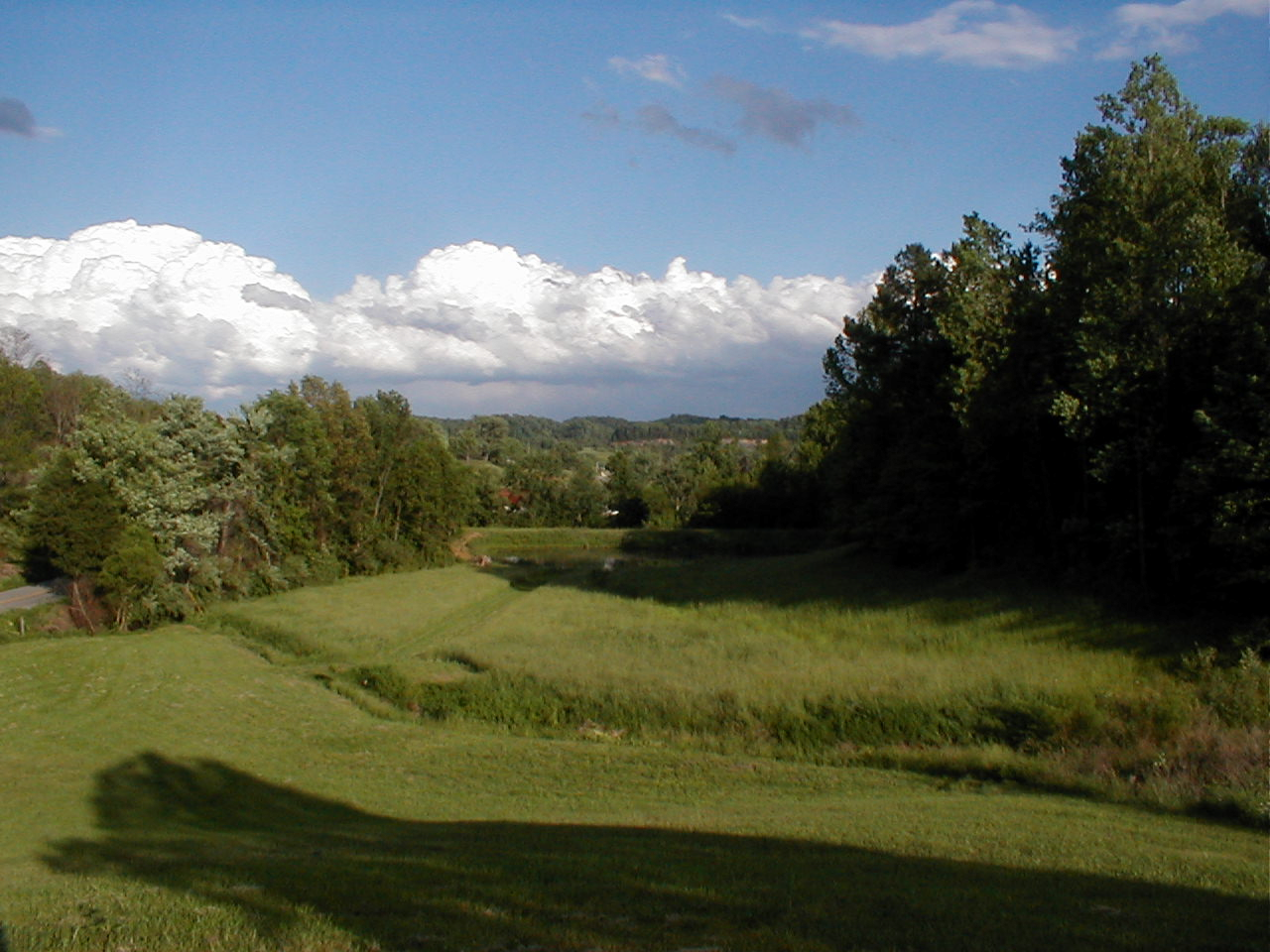}\\
Depths 8 and 7.
  \includegraphics[width=4.7in]{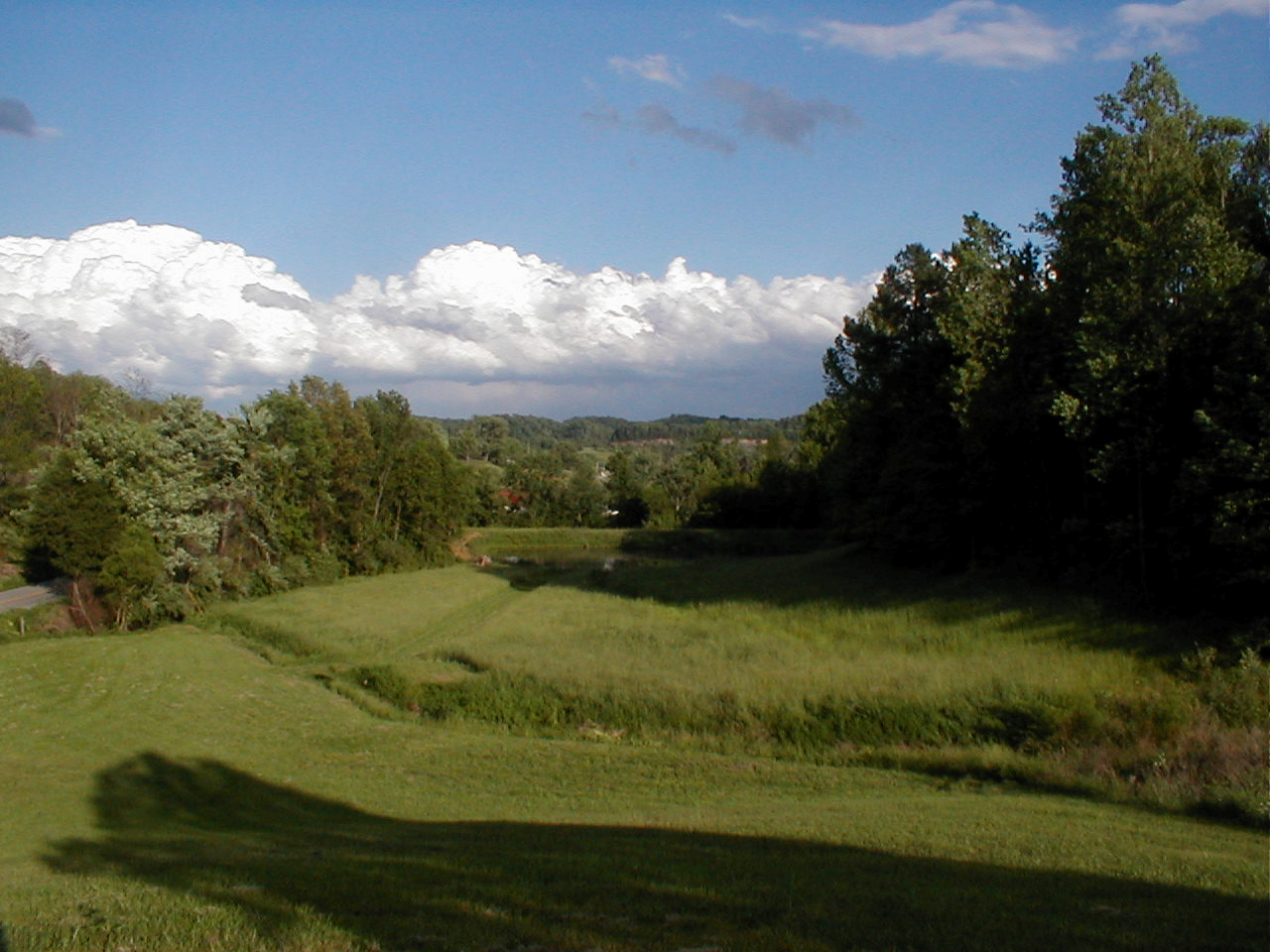}\\
\bigskip
  \includegraphics[width=4.7in]{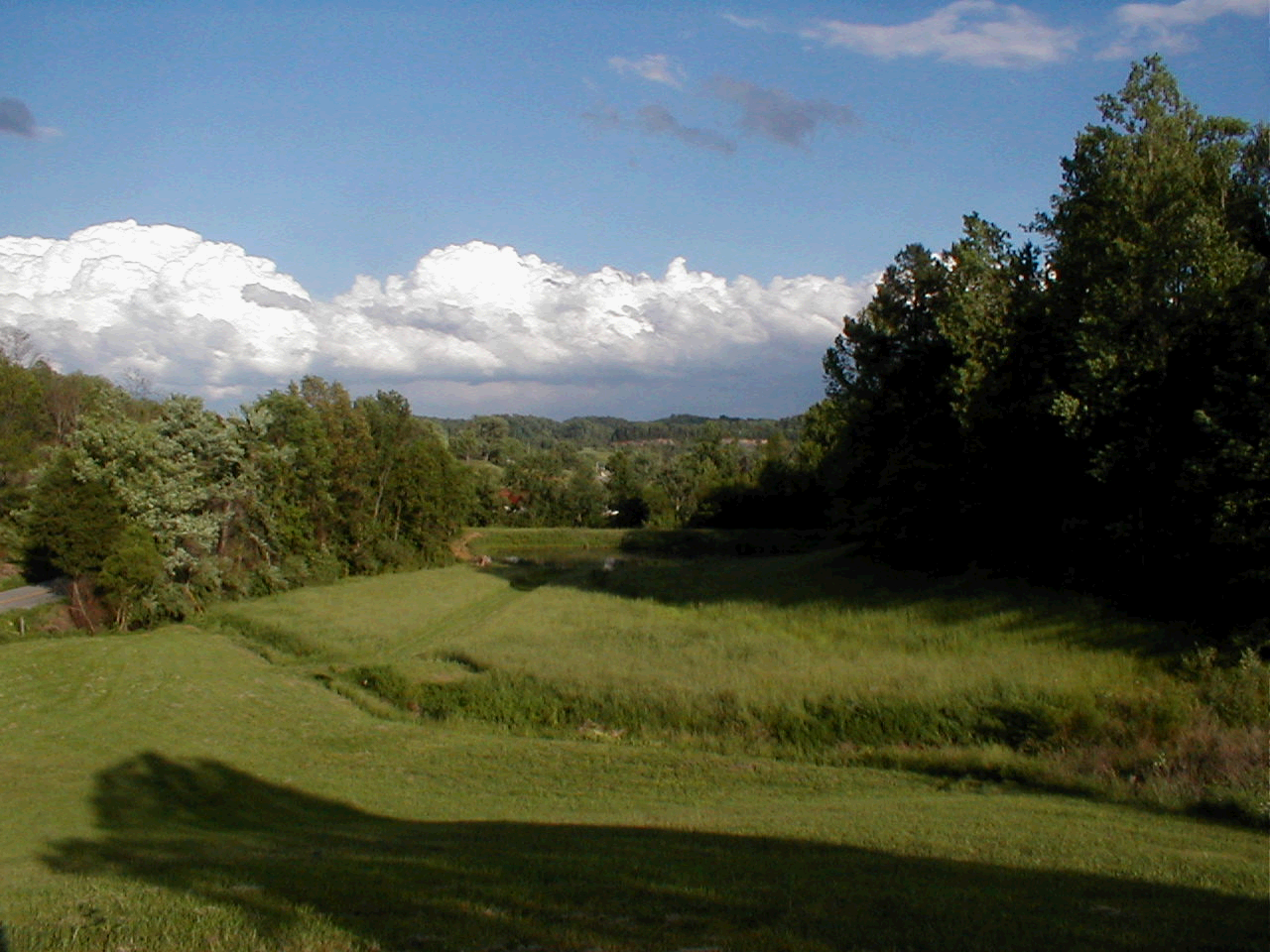}\\
Depths 6 and 5.
  \includegraphics[width=4.7in]{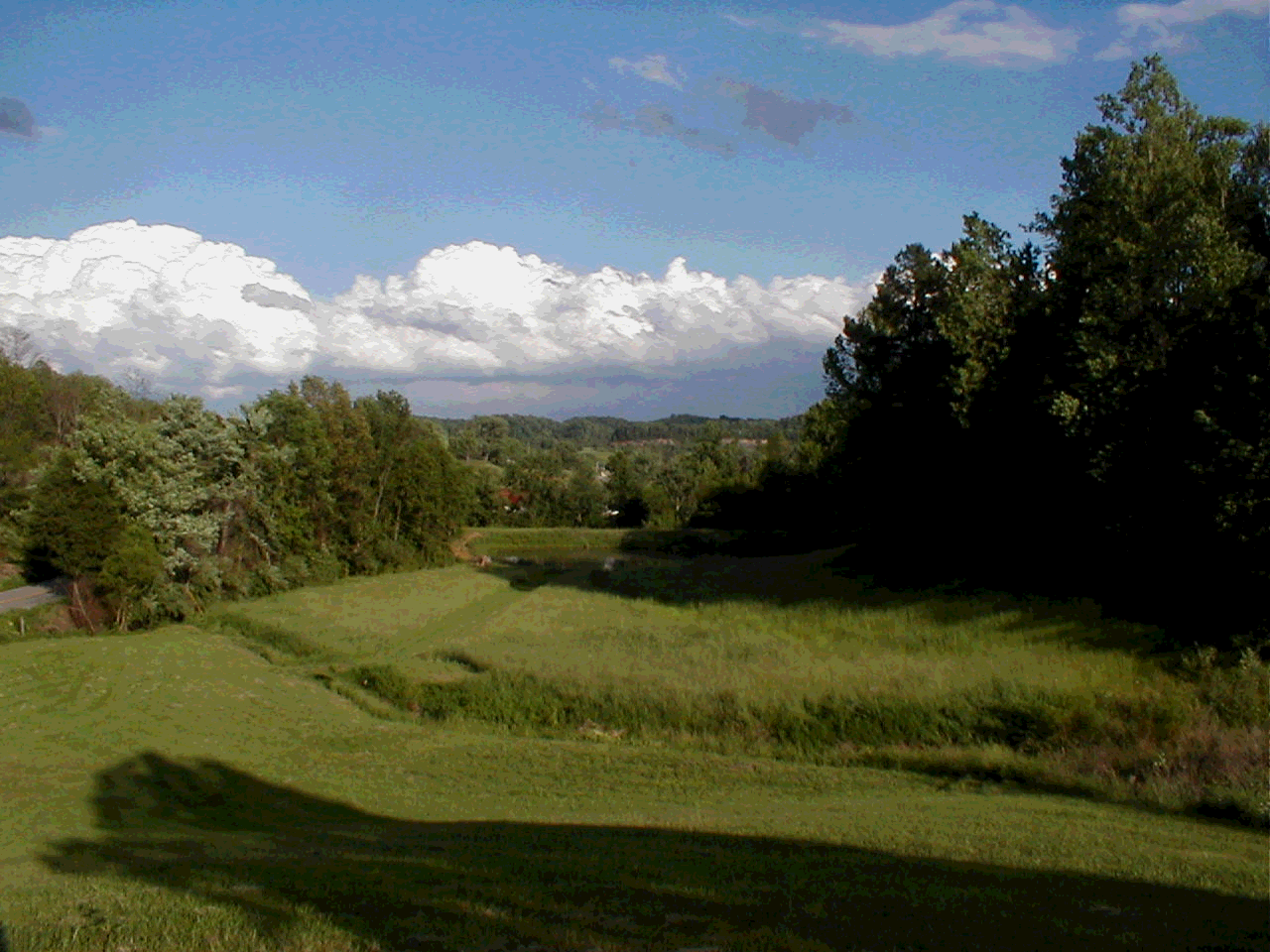}\\
\bigskip
  \includegraphics[width=4.7in]{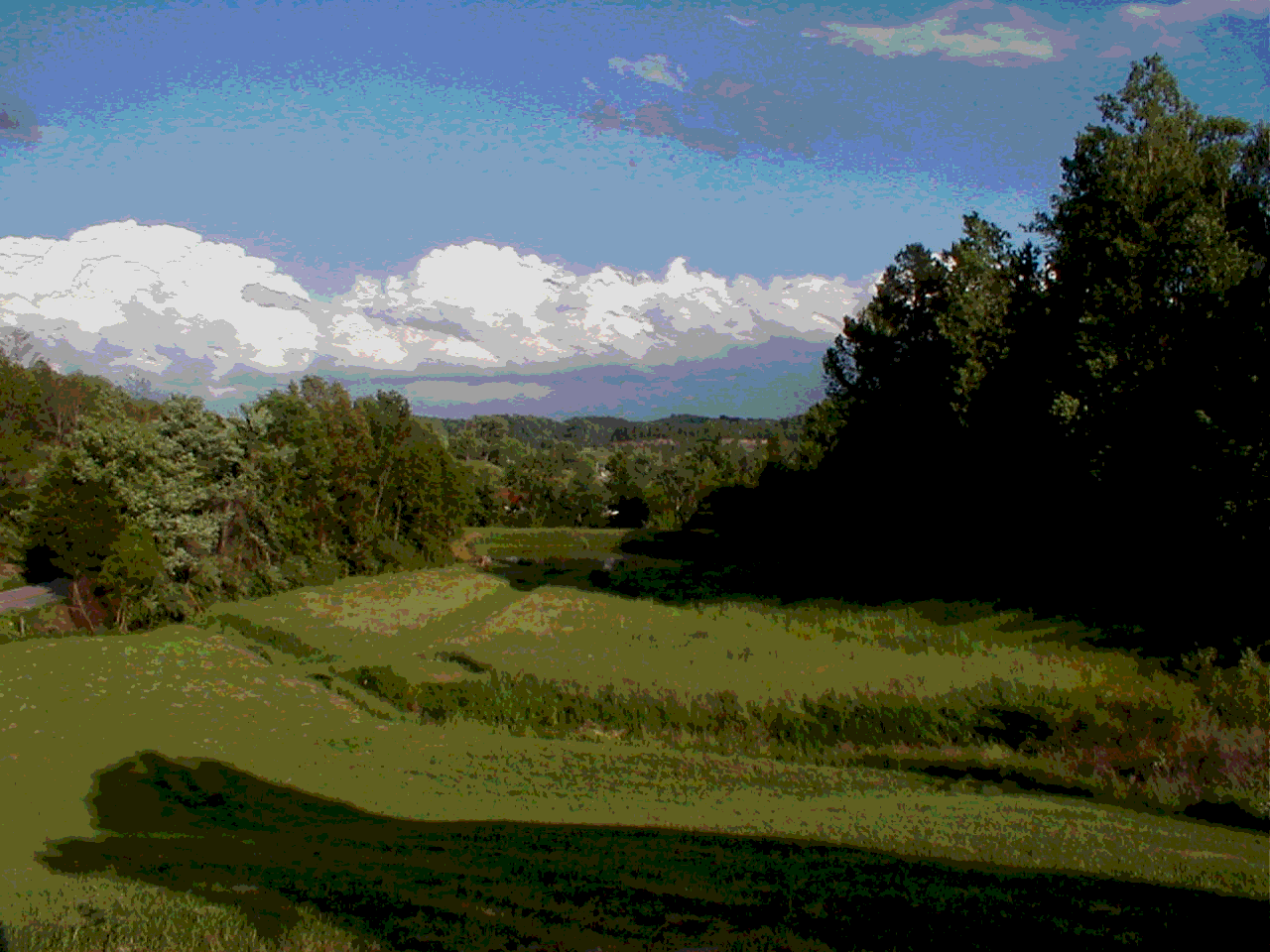}\\
Depths 4 and 3.
  \includegraphics[width=4.7in]{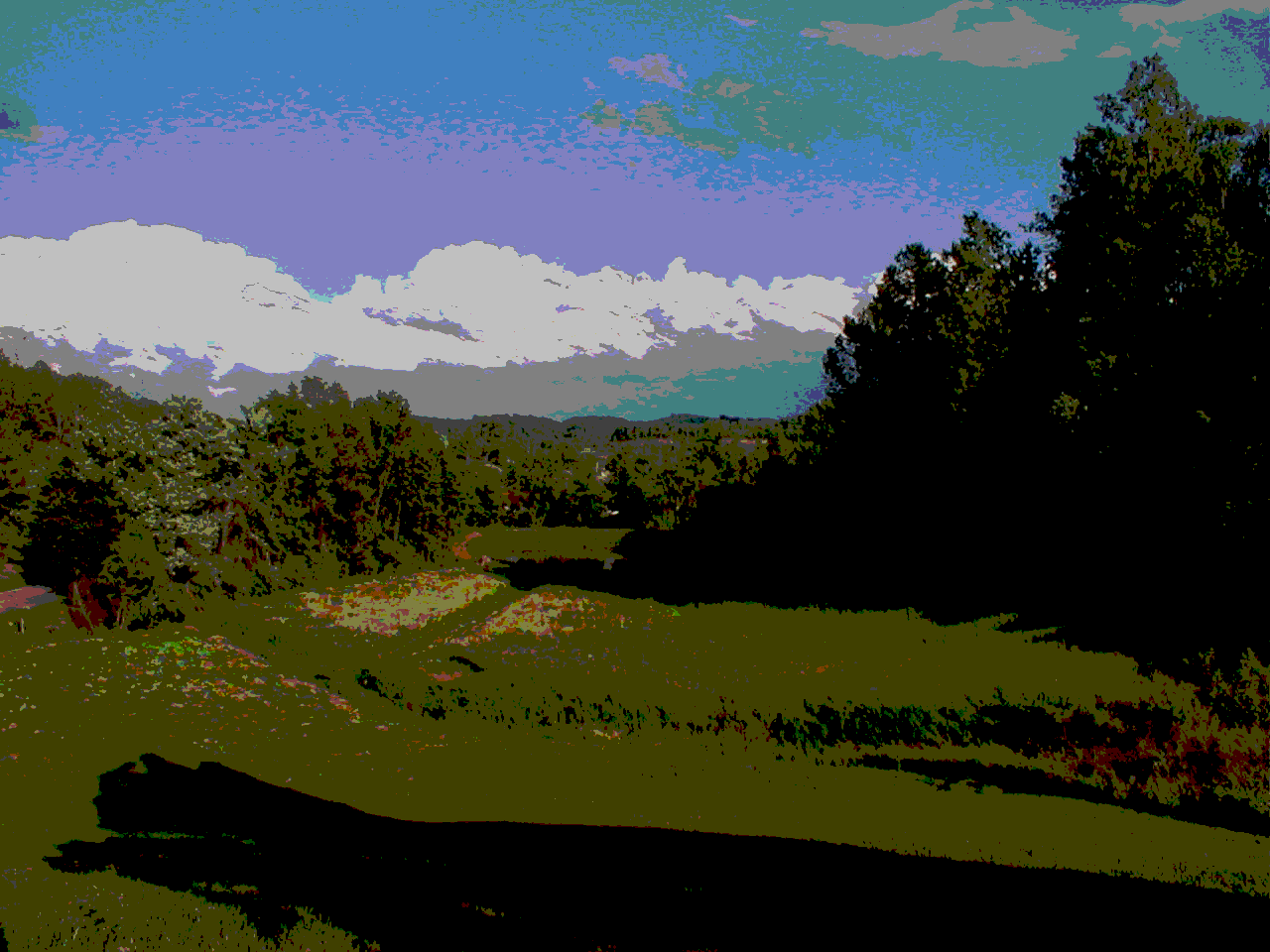}\\
\bigskip
  \includegraphics[width=4.7in]{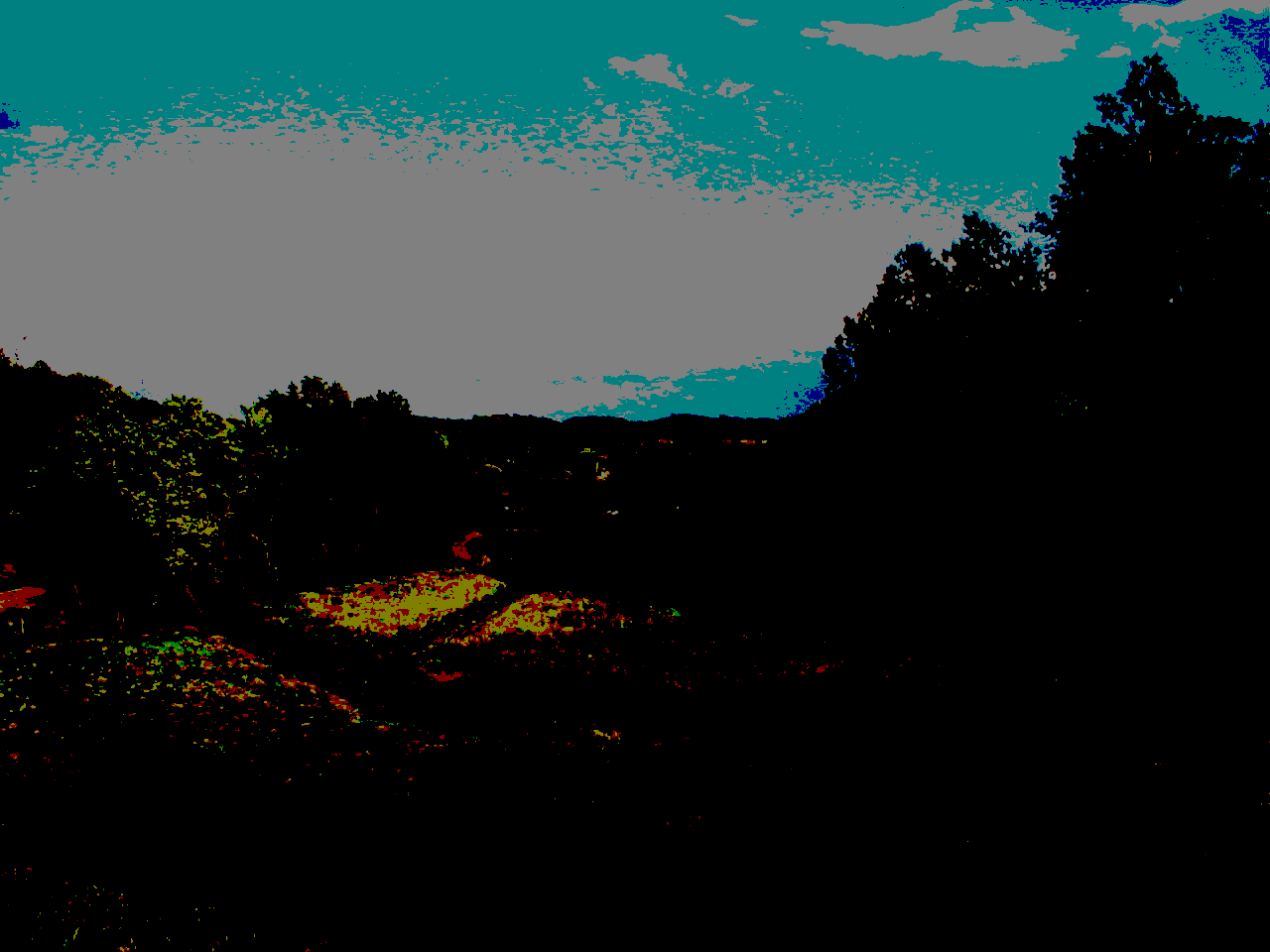}\\
Depths 2 and 1.
\end{center}

Figures 4-11: Signals derived from three independent channels of a color photograph and truncated at depth n=8 through n=1, from upper left to lower right, accordingly.  In this case, the first-order critical depth of the green channel is 4, in the third row at left, however, the first-order depth of the red and blue channels are 3, in the third row at right.  Again, note that the critical depth separates complex, noisy patterns from simple, cartoon-like ones.
\begin{center}
  \includegraphics[width=3.5in]{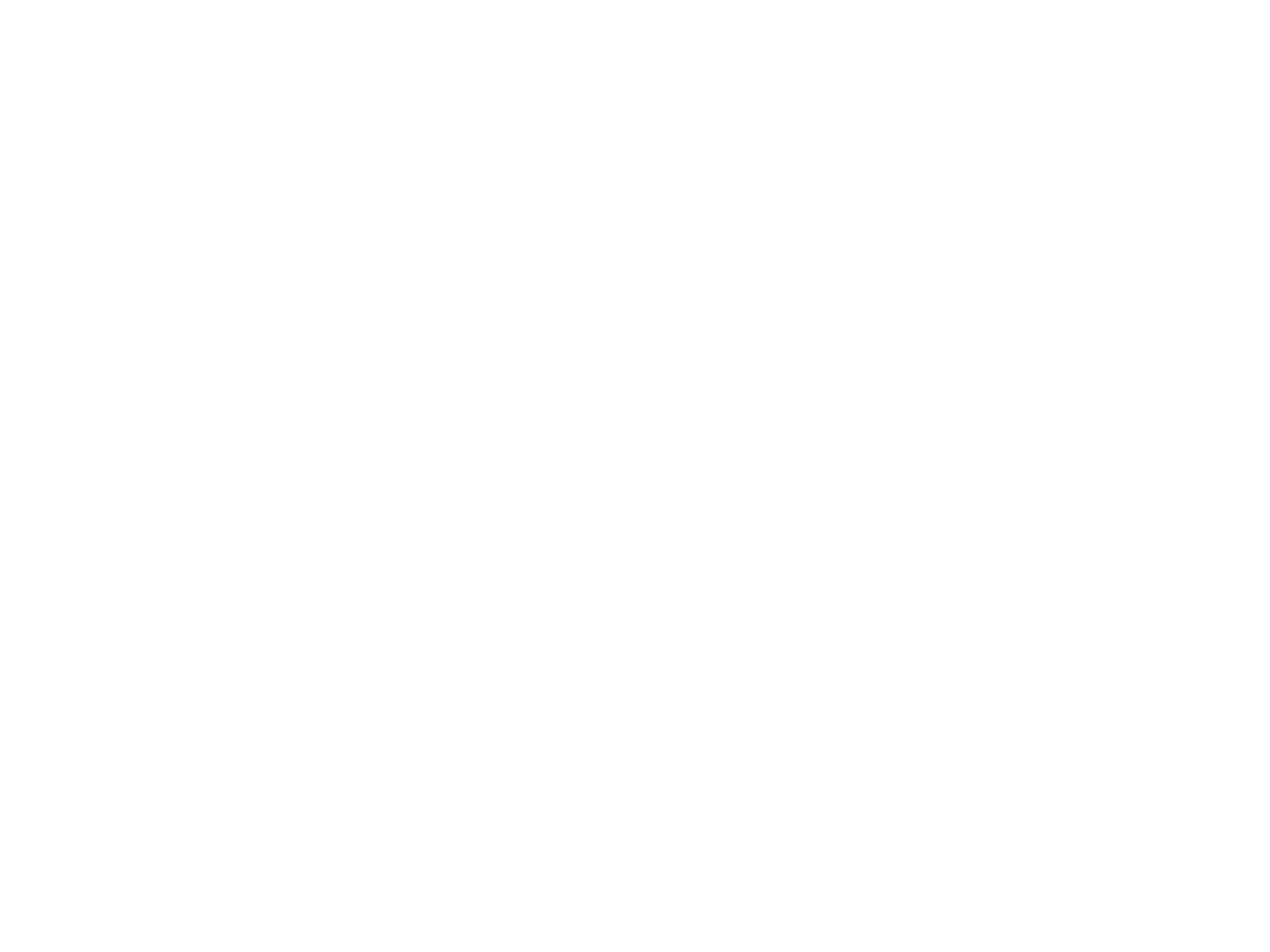}\\
  \includegraphics[width=4.7in]{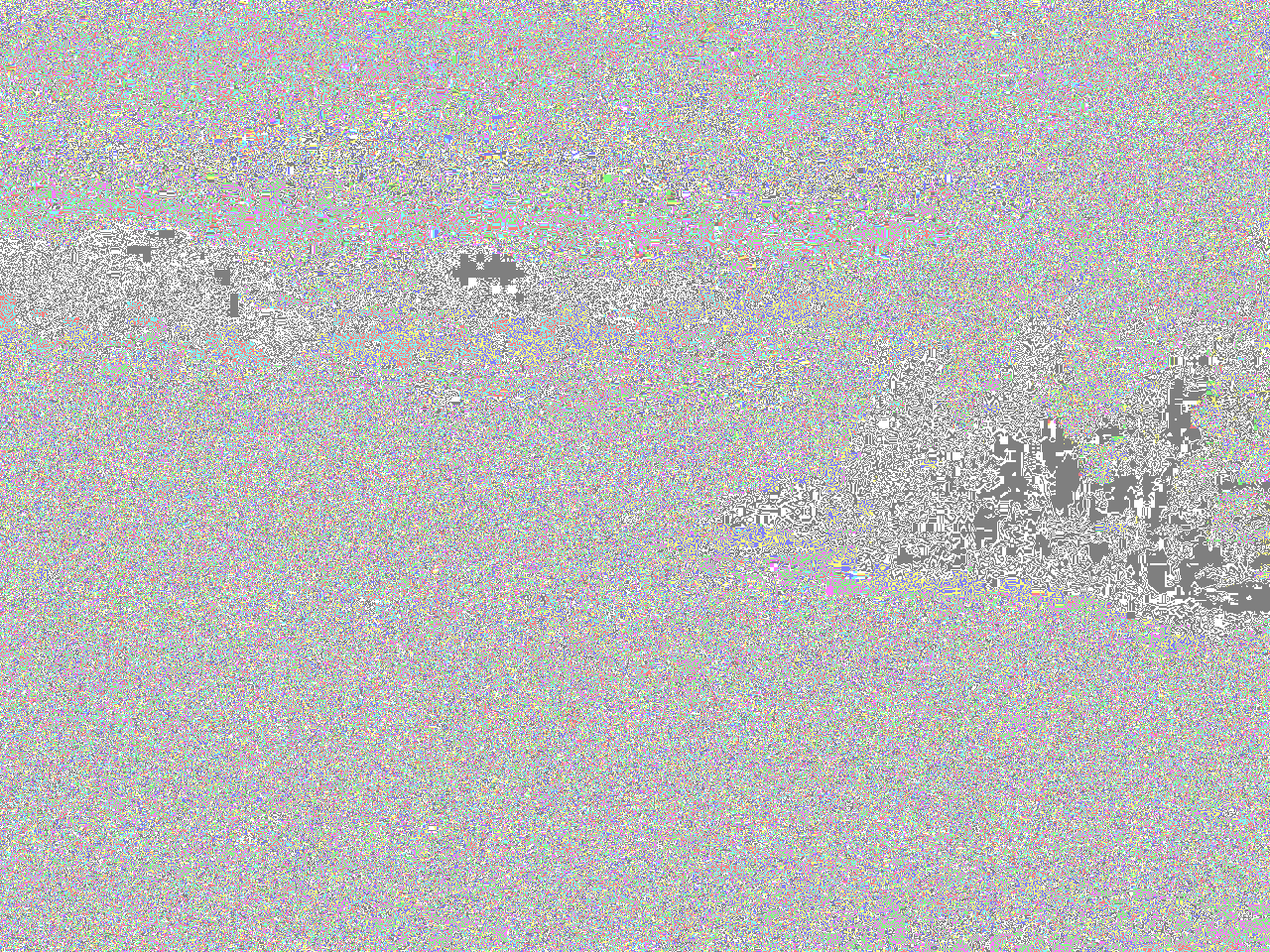}\\
Depths 8 (solid white) and 7.
  \includegraphics[width=4.7in]{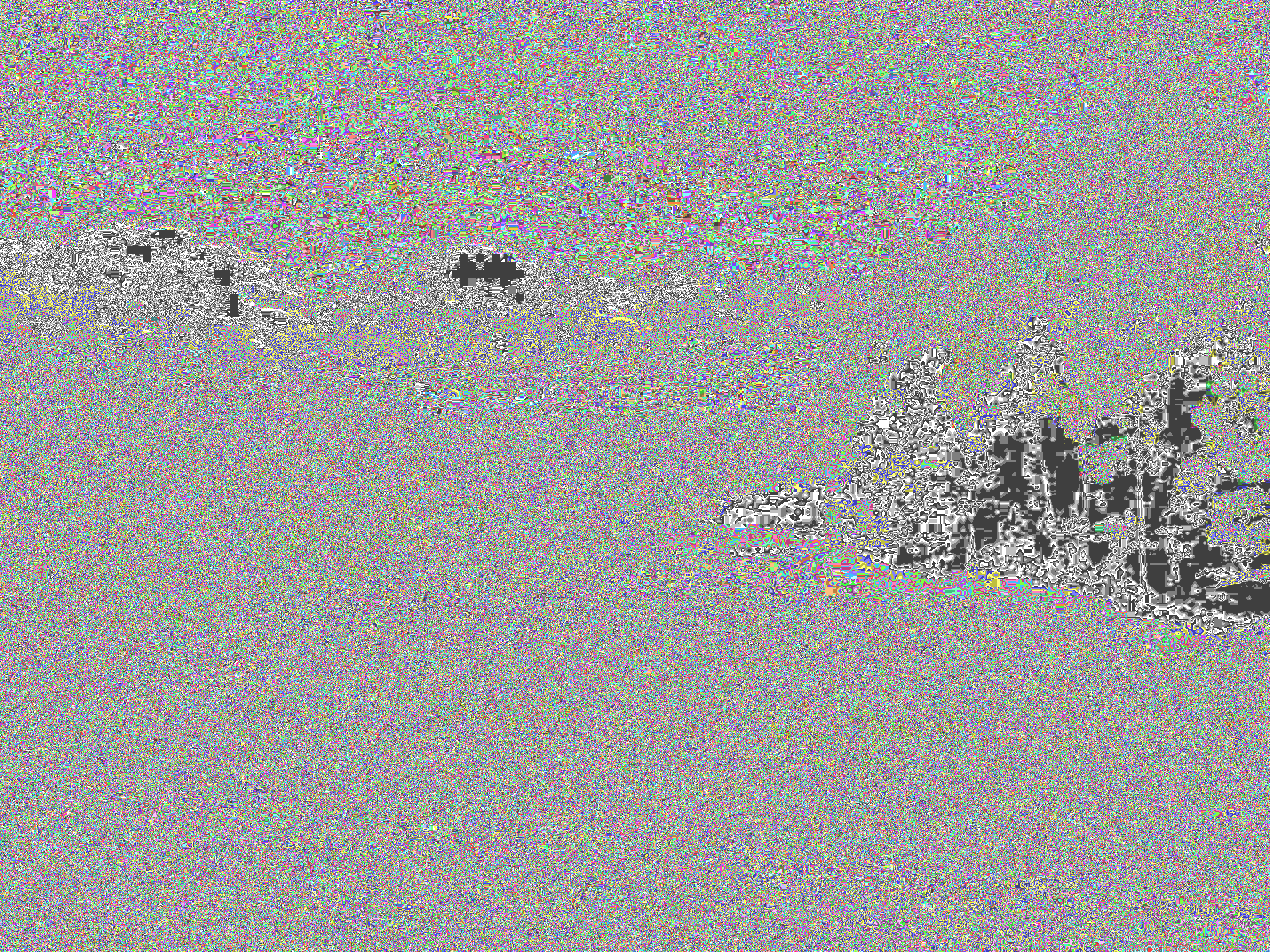}\\
\bigskip
  \includegraphics[width=4.7in]{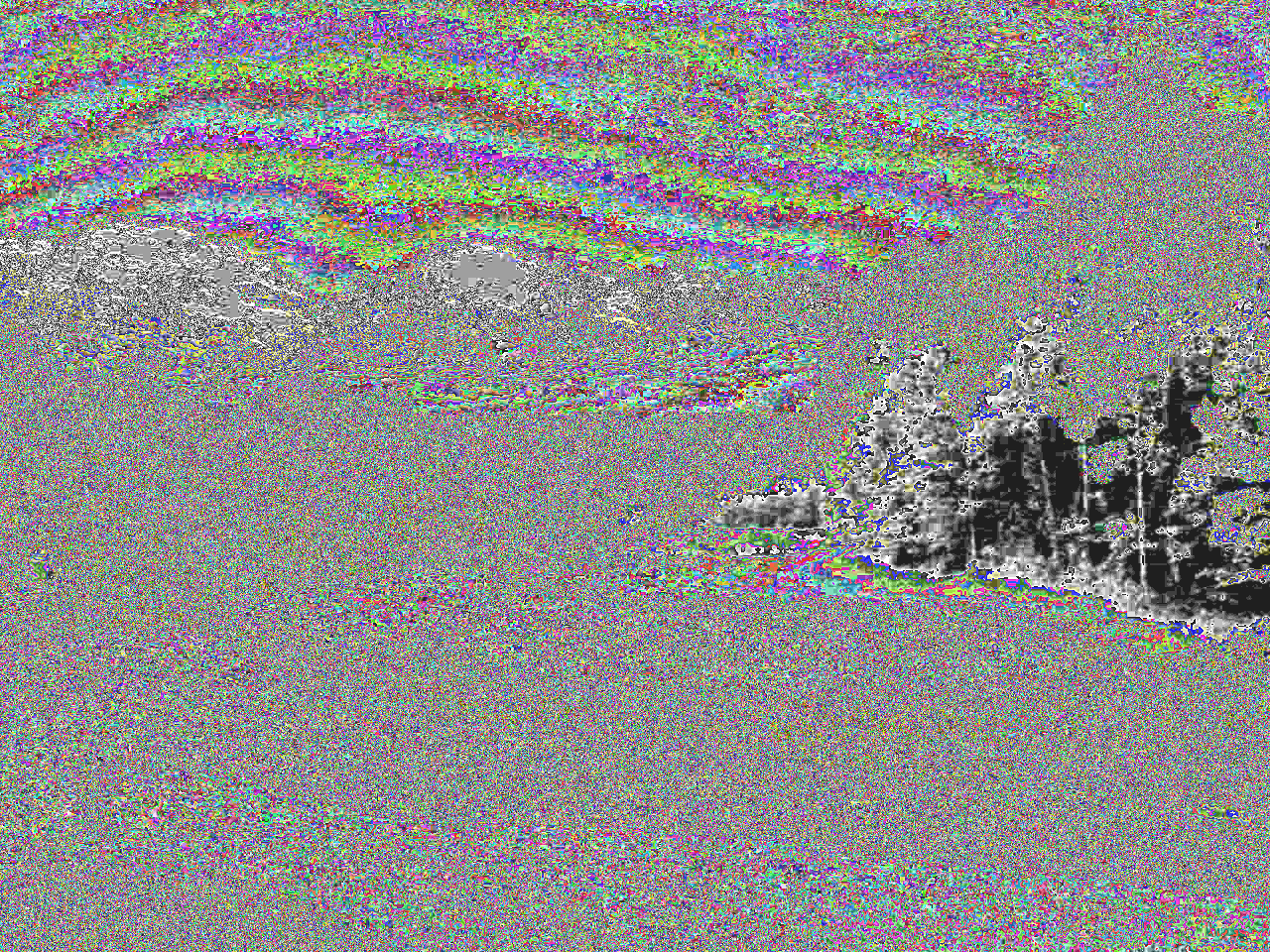}\\
Depths 6 and 5.
  \includegraphics[width=4.7in]{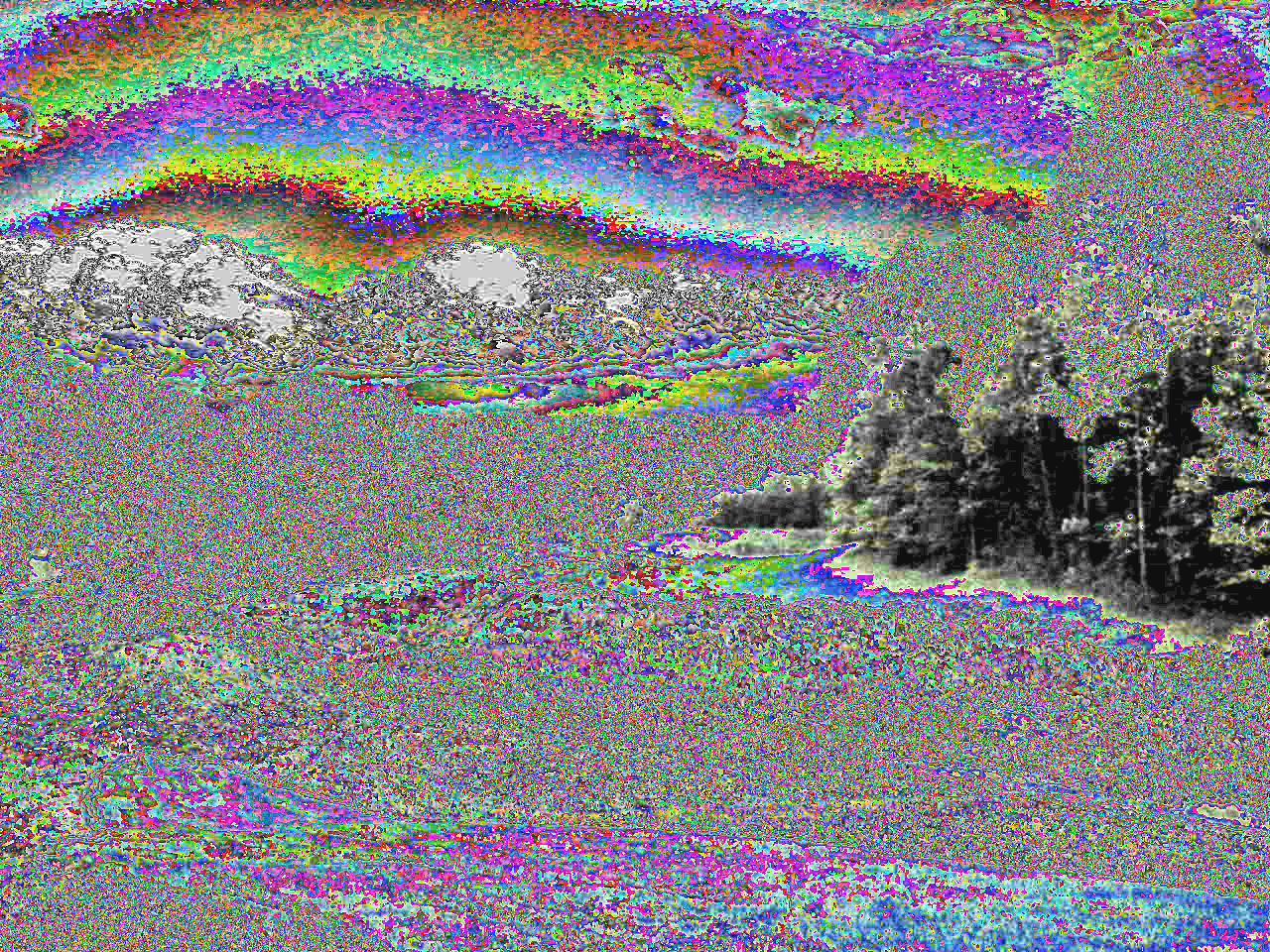}\\
\bigskip
  \includegraphics[width=4.7in]{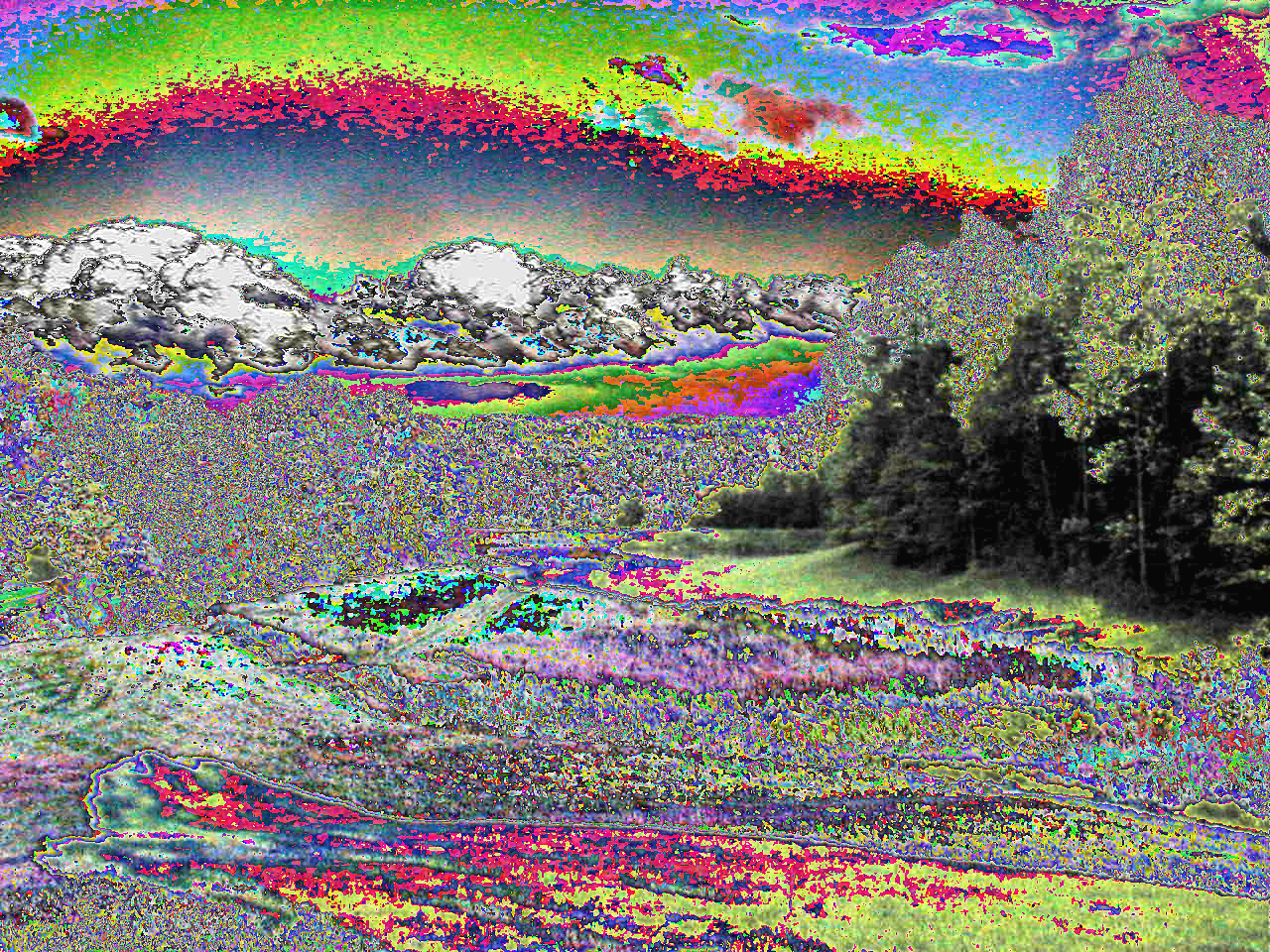}\\
Depths 4 and 3.
  \includegraphics[width=4.7in]{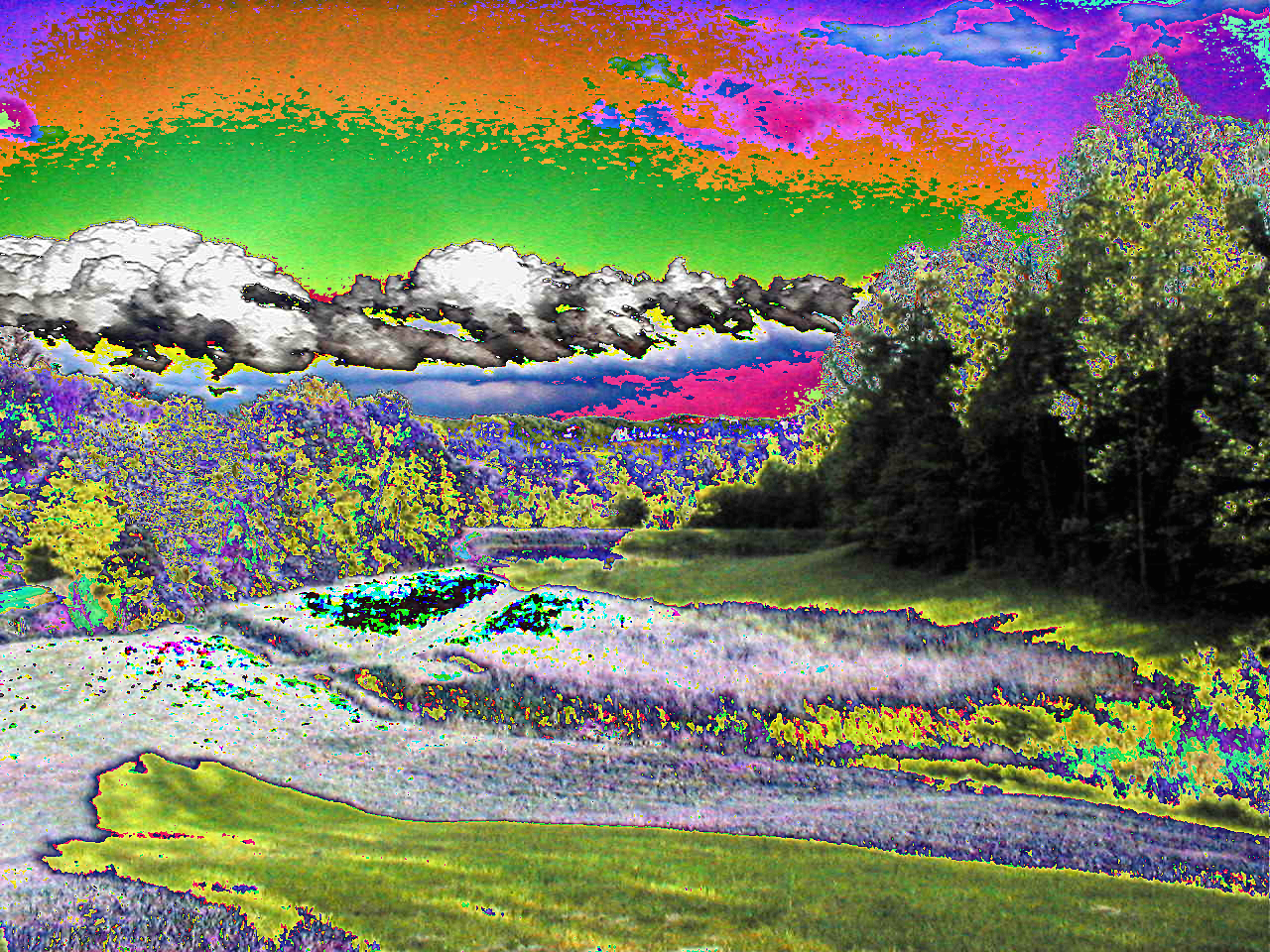}\\
\bigskip
  \includegraphics[width=4.7in]{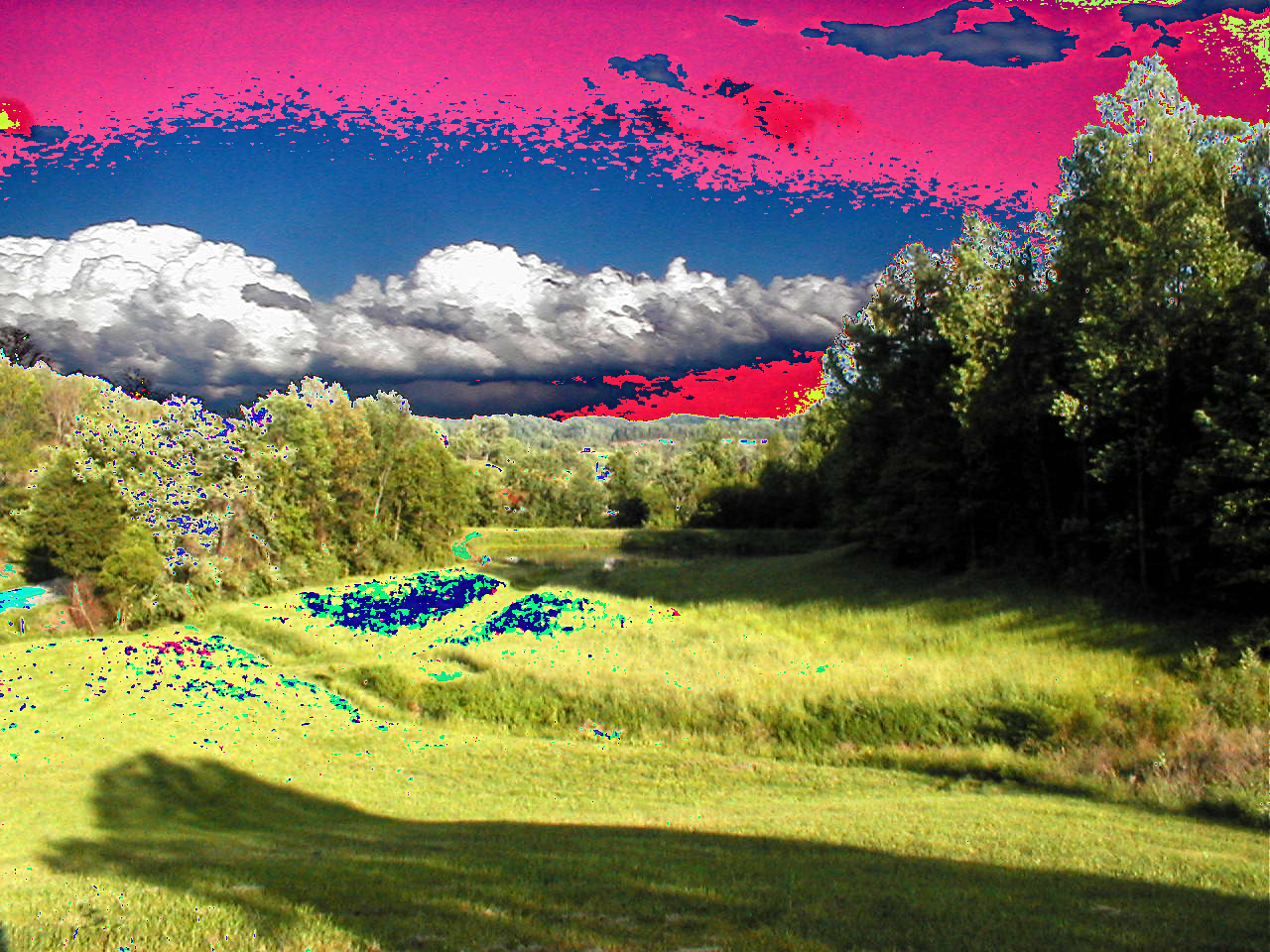}\\
  Depths 2 and 1.
\end{center}

Figures 12-19: The noise truncated from the same signals, n=8 through n=1, normalized to 100\% intensity for easier viewing.  Beyond the first-order critical depth of 4, signal features such as the outlines of trees and clouds creep back into the image, starting with shadows and other darkened regions, suggesting that the bit depth of these regions differs from that of the image as a whole.  The bright blue sky is also at a different bit depth from the overall image, and it is also worth noting that even relatively insignificant bits (n=4 and n=5) retain certain signals.  Due to this sort of inhomogeneity, some cases and applications may benefit from dividing data into blocks (or more generalized regions) and evaluating their critical depth/scale separately.

\bigskip

We are now ready to calculate the critical depth of the color image.  For simplicity, we treat the red, green, and blue channels as being independent tensors.  The image is 1280x960, so each bit of color depth corresponds to 1228800 bits of data.  It had been previously compressed with the JPEG algorithm but this doesn't matter for purposes of illustration - when a more developed approach is evaluated against the mean-squared error of JPEG2000, we will use uncompressed test images.  The $A^n$ are again compressed using gzip until compressing an additional bit of sample depth would increase storage requirements by more than 1228800 bits.  We will start with the red channel, shown below with n on the left and $K(A^n)$ on the right, in bits:
\begin{center}
\begin{tabular}{cc}
  0 & 0 \\
  1 & 188592\\
  2 & 494256 \\
  3 & 1102656 \\
  4 & 1735776 \\
  5 & 3343584 \\
  6 & 4835008 \\
  7 & 7774896 \\
  8 & 8085728 \\
\end{tabular}
\end{center}

Allowing a 25 percent tolerance, the first-order critical depth occurs at n=7 and the second-order critical depth occurs at n=6.  Now we consider the green channel:

\begin{center}
\begin{tabular}{cc}
  0 & 0 \\
  1 & 121616\\
  2 & 427232 \\
  3 & 1038880 \\
  4 & 1643968 \\
  5 & 3161216 \\
  6 & 4606128 \\
  7 & 7374720 \\
  8 & 7764240 \\
\end{tabular}
\end{center}

Again, the first-order critical depth occurs at n=7 and the second-order critical depth occurs at n=6.  Finally, consider the blue channel:
\begin{center}
\begin{tabular}{cc}
  0 & 0 \\
  1 & 51904\\
  2 & 328880 \\
  3 & 839072 \\
  4 & 1655376 \\
  5 & 2961680 \\
  6 & 4423184 \\
  7 & 7124752 \\
  8 & 7902688 \\
\end{tabular}
\end{center}

The result is the same, the first-order critical depth occurs at n=7 and the second-order critical depth occurs at n=6.  We store a critically compressed representation at the second-order critical depth of 6.  Alternately, the complexity of all three channels can be considered simultaneously by compressing the entire image.
\bigskip

Having determined the critical points and stored the associated signal, we are ready to compress its residual noise function.  The critical noise function is compressed using a JPEG algorithm to 134,732 bytes.  The critical signal, compressed via gzip with all three channels in one file, occupies 1,495,987 bytes, which saves 237,053 bytes over compressing the channels separately.  Combined, the critical signal and its residual noise occupy 1,630,719 bytes.  The original image occupies 3,686,400 bytes.  As such, this simplistic implementation of critical compression leads to a compression ratio of just 2.26:1, however, it is essentially indistinguishable (its mean-squared error is just 3.09) from the original image, as may be verified below.
\pagebreak

\begin{center}
  \includegraphics[width=4.7in]{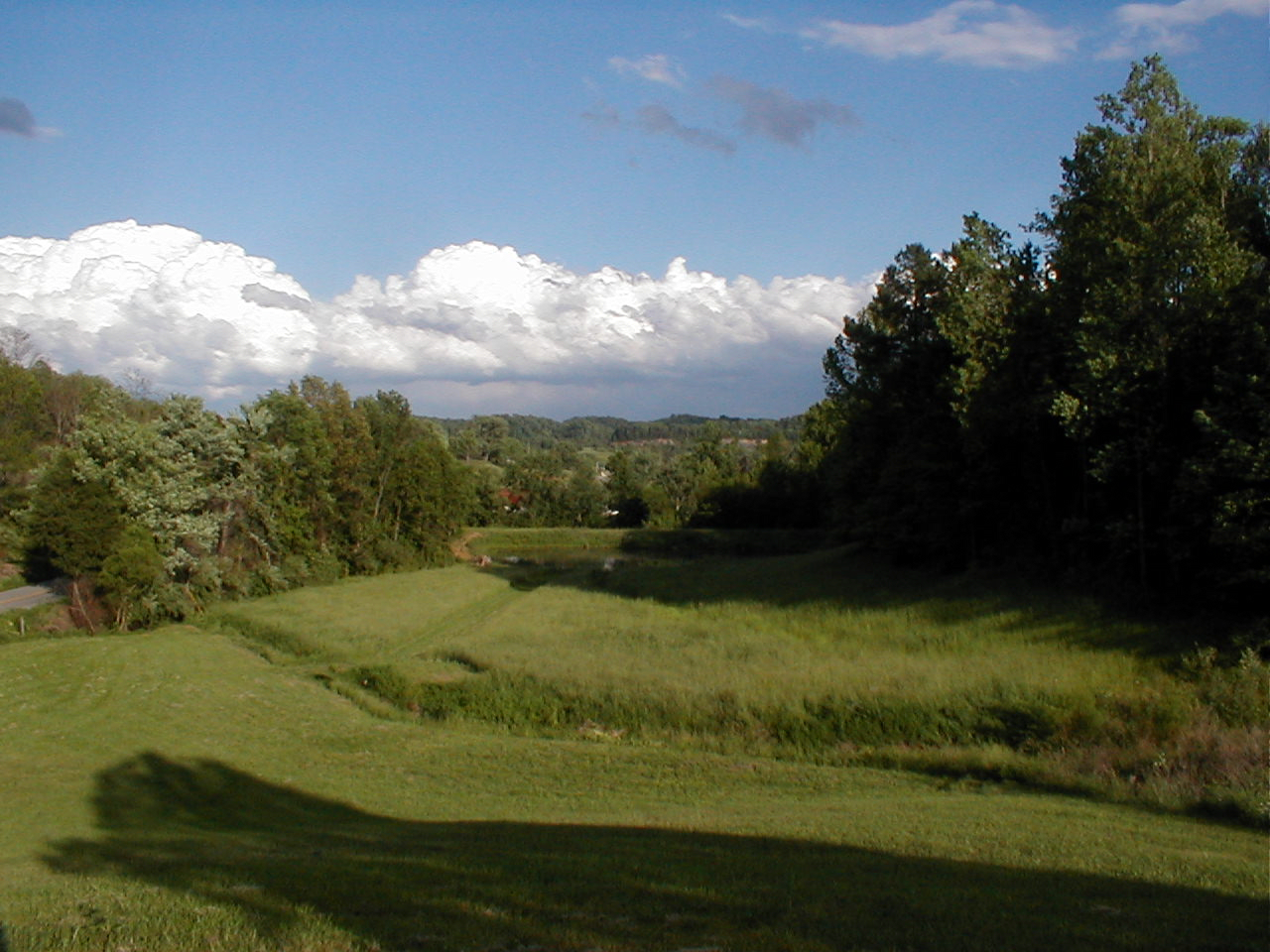}\\
  \smallskip
  \includegraphics[width=4.7in]{fieldcolorsignal8}
\end{center}

Figures 20 and 21: The critically compressed and decompressed image, above, and the original image, below.

\bigskip

The gzip algorithm used to estimate complexities to this point is typical of most lossless data compression in use today, being optimized for speed rather than the highest possible compression ratios.  Naturally, slower but more effective lossless compression will produce estimates closer to the true complexities, leading to better compression and improved inference.

In order to obtain highly compressed lossless codes, we now switch to another algorithm, PAQ8l, which trains and arithmetically codes a neural net which makes linear combinations of nearby pixels or samples.  In this case we encode all three channels of the signal at each bit depth n into a file and apply PAQ using its maximum compression.  In this case, since there are 3 channels at 1280x960, each bit of channel depth constitutes 1,228,800 bits of raw data.  The corresponding estimates of $K(A^n)$, in bits, are in the right column below:
\begin{center}
\begin{tabular}{cc}
  0 & 14552\\
  1 & 168912\\
  2 & 328880 \\
  3 & 1279152 \\
  4 & 2397944 \\
  5 & 4015184 \\
  6 & 6332896 \\
  7 & 9211368 \\
  8 & 11844400 \\
\end{tabular}
\end{center}

To illustrate why Kolmogorov's criterion of randomness does not produce a minimal sufficient statistic in this case, we present the derivative (slope) of complexity with respect to literal description length. In this case, the slope (which is simply $\partial_{t} K(A)$ normalized by $length(A)$ to $[0,1]$) never reaches 1 since even the least significant bits compress by a third or more.  This indicates correlation between the three channels.  The slopes are, ascending in n:
\begin{center}
\begin{tabular}{c}
  n/a \\
  0.0434 \\
  0.0893 \\
   0.2578\\
   0.3035\\
   0.4387\\
   0.6287\\
   0.7808\\
   0.7143
\end{tabular}
\end{center}

This exhibits behavior typical of phase transitions, a sigmoidal form that rises from about 0.04 to over 0.7 as n rises from 0 to 8.  Even so, the complexity estimates are not perfect.  To determine the first-order critical depth, we treat measurements within a $25\%$ tolerance of the maximum value of 0.78084 - measurements greater than 0.58563 - as elements of the maximal set.  This criterion is satisfied at depths 6, 7, and 8, so we select the minimum complexity from this set.  Hence, the first-order critical depth is 6.

By considering the second partial derivative of complexity, the slope of the slope function, using a central difference approximation, we may determine the critical point at which the phase transition proceeds most rapidly.  Since the phase transition is relatively broad, one might expect several points to be close to the maximum rate.  This is the case, as may be seen below:

\begin{center}
\begin{tabular}{c}
  n/a \\
   0.0458\\
   0.1686\\
   0.0471\\
   0.1352\\
   0.1900\\
   0.1521\\
   -0.0666\\
   n/a
\end{tabular}
\end{center}

The maximum (0.1900) occurs at depth n=5.  If we use $25\%$ tolerance, as before, then the cutoff for the maximum set becomes 0.1425075, so its members occur at n=3, n=5, and n=6.  The least complex element in this maximal set determines the second-order critical depth, which in this case is n=3.  This differs significantly from the result obtained using the lower compression levels of gzip, demonstrating the importance of high-quality lossy codes in a critical compression scheme.

We encode the residual noise associated with depth 3 using the JPEG2000 algorithm, which is generally regarded as being slower but higher in quality than JPEG.  For the purposes of constructing critical codes, JPEG2000 has superior properties of convergence to the original image, as compared to JPEG.  Using the Jasper library to compress the noise function at 50:1, the length of our improved depth 3 second-order critical representation is shown below improves to 258,061, while its mean-squared error is about 100, which is typical of broadcast images.  The image is shown below, with the original for comparison.
\pagebreak

\begin{center}
  \includegraphics[width=4.7in]{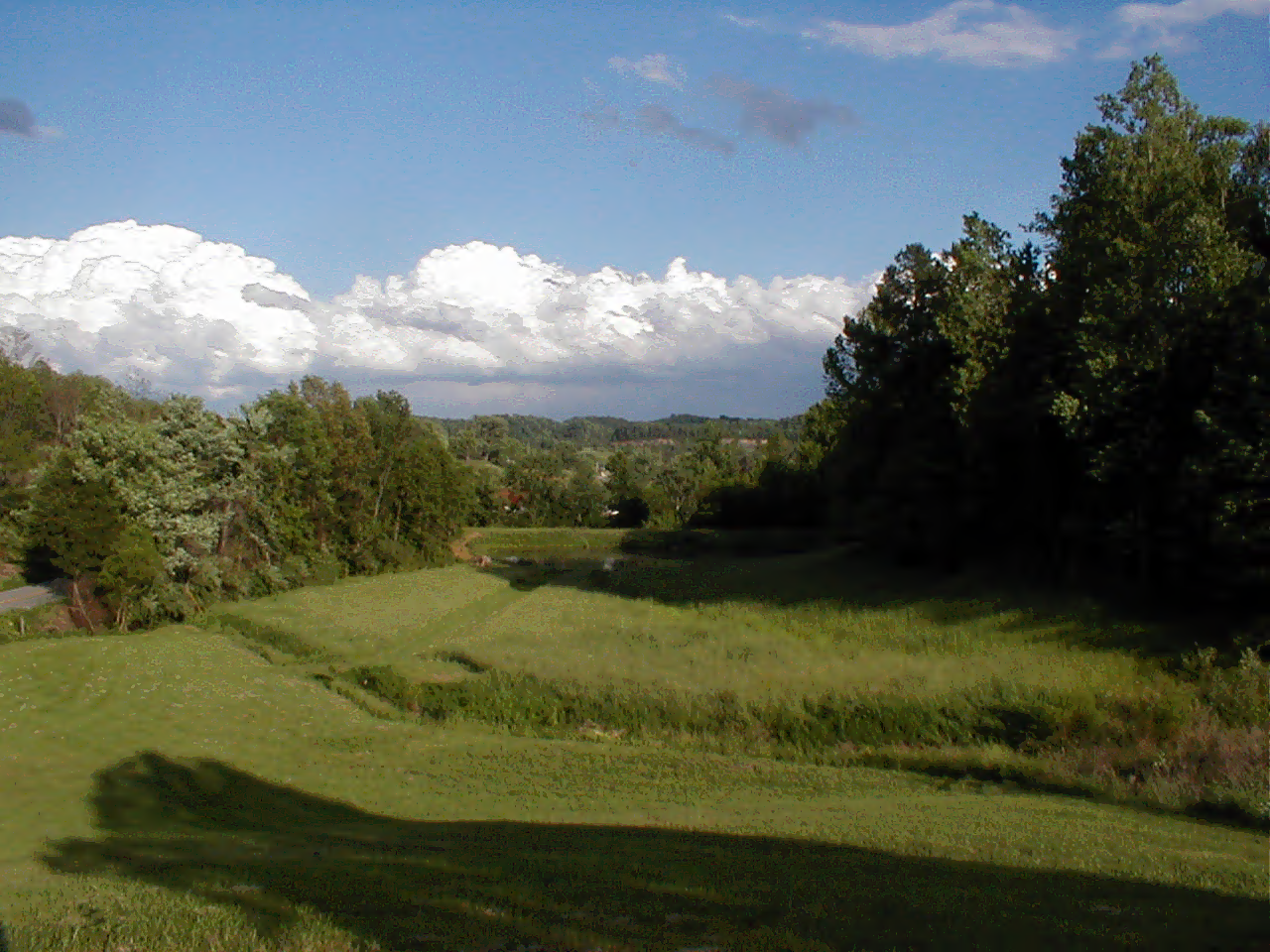}\\
  \smallskip
  \includegraphics[width=4.7in]{fieldcolorsignal8}
\end{center}

Figures 22 and 23:  Improved data compression leads to a better estimate of the second-order critical depth and a more economical representation.  The second order critically compressed and decompressed image (at depth 3) is above and the original image is below.

\bigskip

This simple example shows that critical codes are competitive with the prevailing algorithms for image compression.  In the appendix, we will make a more rigorous determination of the mean squared errors of a variety of standard, uncompressed test images, showing that two part codes actually outperform JPEG2000 for many types of images under this metric.  For images which are not traditional color photographs, the advantage of a critical representation is often significant, demonstrating that two-part codes could be utilized to improve the performance of a wide variety of applications.

\subsection{Example of Video Compression}

We will briefly demonstrate the compression of digital video using 30 uncompressed frames from the movie 'Elephant Dream', whose masters are freely available.  This movie was rendered using computer-generated 3-D graphics, and as a result it has low algorithmic complexity as compared to typical real-world photographic sources.  The frames were rendered using the \emph{Blender} program at 1920x1080 resolution in 24-bit color.

Since video compression proceeds in the same manner as image compression, the only difference being the extra dimension, we may simply interlace the 30 frames into one large (1920x32400) image and perform two-part coding as described previously.  Scanline m of image n is mapped to scanline $30m+n$ of the interlaced image.  In theory, this does negatively impact the compression performance, since each pixel has fewer neighbors, but we will see that the result nonetheless seems to outperform existing video standards, execution speed notwithstanding.

One slight caveat is the size of the interlaced image.  The lossless data compressor (PAQ) eventually compresses the interlaced image, but lossy coding (Jasper JPEG2000) failed for the full noise function due to the image size.  The noise functions were split into two vertical partitions, compressed separately using JPEG2000 coding, decompressed, and stitched back together before being recombined with the lossless signal.  A video of arbitrary duration can be encoded by partitioning its content into blocks in this manner.

Naturally, the division of a video into blocks should proceed such that more compressible content is grouped together.  Lossless compression (especially at a reduced bit depth) is ideal for representing the redundancy that naturally arises from lack of motion in a video sequence.  In an adaptive blocking scheme, the duration of a block takes advantage of this fact, extending farther in time for more redundant scenes, which often have less motion, and shortening the block when relatively incompressible content is reached.  Often, sharp decreases in compressibility along the time axis correspond to rapid change in content such as motion or a scene transition.

We plot the results of critically compressing the interlaced image below.  Each line represents various levels of lossy coding corresponding to some particular signal depth.  The first five depths are shown.  Usually, lower depths have lower complexity and more error than higher depths, but this is not always the case.  Since the video is computer-generated and has relatively little motion, its leading bits are highly compressible.  Depths 1-4 have lossy compression at levels of 400:1, 200:1, and 100:1.  Higher depths have less significant noise functions, so depth 5, the lowest line, has undergone lossy coding at rates of 1000:1 and 500:1.  Since the depth five codes have the desired quality level, subsequent bit depths were not analyzed.

\begin{center}
% GNUPLOT: LaTeX picture
\setlength{\unitlength}{0.240900pt}
\ifx\plotpoint\undefined\newsavebox{\plotpoint}\fi
\sbox{\plotpoint}{\rule[-0.200pt]{0.400pt}{0.400pt}}%
\begin{picture}(1500,900)(0,0)
\sbox{\plotpoint}{\rule[-0.200pt]{0.400pt}{0.400pt}}%
\put(191.0,131.0){\rule[-0.200pt]{4.818pt}{0.400pt}}
\put(171,131){\makebox(0,0)[r]{ 10}}
\put(1419.0,131.0){\rule[-0.200pt]{4.818pt}{0.400pt}}
\put(191.0,228.0){\rule[-0.200pt]{2.409pt}{0.400pt}}
\put(1429.0,228.0){\rule[-0.200pt]{2.409pt}{0.400pt}}
\put(191.0,285.0){\rule[-0.200pt]{2.409pt}{0.400pt}}
\put(1429.0,285.0){\rule[-0.200pt]{2.409pt}{0.400pt}}
\put(191.0,325.0){\rule[-0.200pt]{2.409pt}{0.400pt}}
\put(1429.0,325.0){\rule[-0.200pt]{2.409pt}{0.400pt}}
\put(191.0,356.0){\rule[-0.200pt]{2.409pt}{0.400pt}}
\put(1429.0,356.0){\rule[-0.200pt]{2.409pt}{0.400pt}}
\put(191.0,382.0){\rule[-0.200pt]{2.409pt}{0.400pt}}
\put(1429.0,382.0){\rule[-0.200pt]{2.409pt}{0.400pt}}
\put(191.0,404.0){\rule[-0.200pt]{2.409pt}{0.400pt}}
\put(1429.0,404.0){\rule[-0.200pt]{2.409pt}{0.400pt}}
\put(191.0,422.0){\rule[-0.200pt]{2.409pt}{0.400pt}}
\put(1429.0,422.0){\rule[-0.200pt]{2.409pt}{0.400pt}}
\put(191.0,439.0){\rule[-0.200pt]{2.409pt}{0.400pt}}
\put(1429.0,439.0){\rule[-0.200pt]{2.409pt}{0.400pt}}
\put(191.0,454.0){\rule[-0.200pt]{4.818pt}{0.400pt}}
\put(171,454){\makebox(0,0)[r]{ 100}}
\put(1419.0,454.0){\rule[-0.200pt]{4.818pt}{0.400pt}}
\put(191.0,551.0){\rule[-0.200pt]{2.409pt}{0.400pt}}
\put(1429.0,551.0){\rule[-0.200pt]{2.409pt}{0.400pt}}
\put(191.0,607.0){\rule[-0.200pt]{2.409pt}{0.400pt}}
\put(1429.0,607.0){\rule[-0.200pt]{2.409pt}{0.400pt}}
\put(191.0,648.0){\rule[-0.200pt]{2.409pt}{0.400pt}}
\put(1429.0,648.0){\rule[-0.200pt]{2.409pt}{0.400pt}}
\put(191.0,679.0){\rule[-0.200pt]{2.409pt}{0.400pt}}
\put(1429.0,679.0){\rule[-0.200pt]{2.409pt}{0.400pt}}
\put(191.0,704.0){\rule[-0.200pt]{2.409pt}{0.400pt}}
\put(1429.0,704.0){\rule[-0.200pt]{2.409pt}{0.400pt}}
\put(191.0,726.0){\rule[-0.200pt]{2.409pt}{0.400pt}}
\put(1429.0,726.0){\rule[-0.200pt]{2.409pt}{0.400pt}}
\put(191.0,745.0){\rule[-0.200pt]{2.409pt}{0.400pt}}
\put(1429.0,745.0){\rule[-0.200pt]{2.409pt}{0.400pt}}
\put(191.0,761.0){\rule[-0.200pt]{2.409pt}{0.400pt}}
\put(1429.0,761.0){\rule[-0.200pt]{2.409pt}{0.400pt}}
\put(191.0,776.0){\rule[-0.200pt]{4.818pt}{0.400pt}}
\put(171,776){\makebox(0,0)[r]{ 1000}}
\put(1419.0,776.0){\rule[-0.200pt]{4.818pt}{0.400pt}}
\put(191.0,131.0){\rule[-0.200pt]{0.400pt}{4.818pt}}
\put(191,90){\makebox(0,0){ 100000}}
\put(191.0,756.0){\rule[-0.200pt]{0.400pt}{4.818pt}}
\put(379.0,131.0){\rule[-0.200pt]{0.400pt}{2.409pt}}
\put(379.0,766.0){\rule[-0.200pt]{0.400pt}{2.409pt}}
\put(489.0,131.0){\rule[-0.200pt]{0.400pt}{2.409pt}}
\put(489.0,766.0){\rule[-0.200pt]{0.400pt}{2.409pt}}
\put(567.0,131.0){\rule[-0.200pt]{0.400pt}{2.409pt}}
\put(567.0,766.0){\rule[-0.200pt]{0.400pt}{2.409pt}}
\put(627.0,131.0){\rule[-0.200pt]{0.400pt}{2.409pt}}
\put(627.0,766.0){\rule[-0.200pt]{0.400pt}{2.409pt}}
\put(677.0,131.0){\rule[-0.200pt]{0.400pt}{2.409pt}}
\put(677.0,766.0){\rule[-0.200pt]{0.400pt}{2.409pt}}
\put(718.0,131.0){\rule[-0.200pt]{0.400pt}{2.409pt}}
\put(718.0,766.0){\rule[-0.200pt]{0.400pt}{2.409pt}}
\put(755.0,131.0){\rule[-0.200pt]{0.400pt}{2.409pt}}
\put(755.0,766.0){\rule[-0.200pt]{0.400pt}{2.409pt}}
\put(786.0,131.0){\rule[-0.200pt]{0.400pt}{2.409pt}}
\put(786.0,766.0){\rule[-0.200pt]{0.400pt}{2.409pt}}
\put(815.0,131.0){\rule[-0.200pt]{0.400pt}{4.818pt}}
\put(815,90){\makebox(0,0){ 1e+006}}
\put(815.0,756.0){\rule[-0.200pt]{0.400pt}{4.818pt}}
\put(1003.0,131.0){\rule[-0.200pt]{0.400pt}{2.409pt}}
\put(1003.0,766.0){\rule[-0.200pt]{0.400pt}{2.409pt}}
\put(1113.0,131.0){\rule[-0.200pt]{0.400pt}{2.409pt}}
\put(1113.0,766.0){\rule[-0.200pt]{0.400pt}{2.409pt}}
\put(1191.0,131.0){\rule[-0.200pt]{0.400pt}{2.409pt}}
\put(1191.0,766.0){\rule[-0.200pt]{0.400pt}{2.409pt}}
\put(1251.0,131.0){\rule[-0.200pt]{0.400pt}{2.409pt}}
\put(1251.0,766.0){\rule[-0.200pt]{0.400pt}{2.409pt}}
\put(1301.0,131.0){\rule[-0.200pt]{0.400pt}{2.409pt}}
\put(1301.0,766.0){\rule[-0.200pt]{0.400pt}{2.409pt}}
\put(1342.0,131.0){\rule[-0.200pt]{0.400pt}{2.409pt}}
\put(1342.0,766.0){\rule[-0.200pt]{0.400pt}{2.409pt}}
\put(1379.0,131.0){\rule[-0.200pt]{0.400pt}{2.409pt}}
\put(1379.0,766.0){\rule[-0.200pt]{0.400pt}{2.409pt}}
\put(1410.0,131.0){\rule[-0.200pt]{0.400pt}{2.409pt}}
\put(1410.0,766.0){\rule[-0.200pt]{0.400pt}{2.409pt}}
\put(1439.0,131.0){\rule[-0.200pt]{0.400pt}{4.818pt}}
\put(1439,90){\makebox(0,0){ 1e+007}}
\put(1439.0,756.0){\rule[-0.200pt]{0.400pt}{4.818pt}}
\put(191.0,131.0){\rule[-0.200pt]{0.400pt}{155.380pt}}
\put(191.0,131.0){\rule[-0.200pt]{300.643pt}{0.400pt}}
\put(1439.0,131.0){\rule[-0.200pt]{0.400pt}{155.380pt}}
\put(191.0,776.0){\rule[-0.200pt]{300.643pt}{0.400pt}}
\put(30,453){\makebox(0,0){\shortstack{Mean\\Squared\\Error}}}
\put(815,29){\makebox(0,0){Compressed Size}}
\put(815,838){\makebox(0,0){Compression Performance, 'Elephant Dream', Frames 1000-1029}}
\put(846,669){\usebox{\plotpoint}}
\multiput(846.58,666.78)(0.499,-0.542){235}{\rule{0.120pt}{0.534pt}}
\multiput(845.17,667.89)(119.000,-127.892){2}{\rule{0.400pt}{0.267pt}}
\multiput(965.58,537.90)(0.499,-0.507){291}{\rule{0.120pt}{0.505pt}}
\multiput(964.17,538.95)(147.000,-147.951){2}{\rule{0.400pt}{0.253pt}}
\put(804,592){\usebox{\plotpoint}}
\multiput(804.00,590.92)(0.795,-0.499){167}{\rule{0.735pt}{0.120pt}}
\multiput(804.00,591.17)(133.474,-85.000){2}{\rule{0.368pt}{0.400pt}}
\multiput(939.58,504.53)(0.499,-0.618){243}{\rule{0.120pt}{0.594pt}}
\multiput(938.17,505.77)(123.000,-150.766){2}{\rule{0.400pt}{0.297pt}}
\sbox{\plotpoint}{\rule[-0.400pt]{0.800pt}{0.800pt}}%
\sbox{\plotpoint}{\rule[-0.200pt]{0.400pt}{0.400pt}}%
\put(840,468){\usebox{\plotpoint}}
\multiput(840.00,466.92)(1.054,-0.499){113}{\rule{0.941pt}{0.120pt}}
\multiput(840.00,467.17)(120.046,-58.000){2}{\rule{0.471pt}{0.400pt}}
\multiput(962.00,408.92)(0.607,-0.499){239}{\rule{0.586pt}{0.120pt}}
\multiput(962.00,409.17)(145.784,-121.000){2}{\rule{0.293pt}{0.400pt}}
\sbox{\plotpoint}{\rule[-0.500pt]{1.000pt}{1.000pt}}%
\sbox{\plotpoint}{\rule[-0.200pt]{0.400pt}{0.400pt}}%
\put(883,315){\usebox{\plotpoint}}
\multiput(883.00,315.58)(0.645,0.499){163}{\rule{0.616pt}{0.120pt}}
\multiput(883.00,314.17)(105.722,83.000){2}{\rule{0.308pt}{0.400pt}}
\multiput(990.58,394.90)(0.499,-0.809){269}{\rule{0.120pt}{0.747pt}}
\multiput(989.17,396.45)(136.000,-218.449){2}{\rule{0.400pt}{0.374pt}}
\sbox{\plotpoint}{\rule[-0.600pt]{1.200pt}{1.200pt}}%
\sbox{\plotpoint}{\rule[-0.200pt]{0.400pt}{0.400pt}}%
\put(860,176){\usebox{\plotpoint}}
\multiput(860.00,174.92)(1.650,-0.494){29}{\rule{1.400pt}{0.119pt}}
\multiput(860.00,175.17)(49.094,-16.000){2}{\rule{0.700pt}{0.400pt}}
\put(191.0,131.0){\rule[-0.200pt]{0.400pt}{155.380pt}}
\put(191.0,131.0){\rule[-0.200pt]{300.643pt}{0.400pt}}
\put(1439.0,131.0){\rule[-0.200pt]{0.400pt}{155.380pt}}
\put(191.0,776.0){\rule[-0.200pt]{300.643pt}{0.400pt}}
\end{picture}
\end{center}

Since the clip tested has only 30 frames, results will not be compared in detail to existing algorithms at this time.  It is worth noting, however, that 1080p video generally needs more than the 1 megabyte per second rate at which this two part code achieves high fidelity.  The best two-part code (depth 5 with 1000:1 lossy coding) is 1,179,977 bytes long, which constitutes a 158:1 compression ratio, yet its mean squared error is only 13.77, making it indistinguishable to the naked eye.  Benchmarking both animated and photographic video content will be the topic of a future study, however, the performance of this simplistic implementation suggests that animation would often be better represented by using a two part code (which could in some cases revert to pure lossless coding) than with conventional transformation-based lossy video codecs.

Extending the method to rank 3 tensors using (for instance) motion JPEG2000 coding would tend to increase the redundancy of local information and hence compression performance.  The underlying lossless code could also be extended to take explicit advantage of rank-3 tensors even though an idealized entropic code would automatically take advantage of this redundancy.

\subsection{Example of Audio Compression}

As mentioned previously, contemporary digital audio recordings are usually sampled above the Nyquist rate associated with human hearing, 44100Hz being common in consumer applications.  This is in contrast to images, where the Nyquist rate would be half of the nanometer-scale wavelength of visible light, beyond the resolution of most traditional optics.  As such, audio signals can be reliably resolved in the frequency domain, whereas most photographs cannot.

As such, to compress audio, we divide the signal into some number of blocks and calculate the spectrum of each block by evaluating its discrete Fourier transform, optionally using the fast Fourier transform if the block size is a power of two.  We then encode a graph of each of these spectra into an image having height one and width equal to the block size and proceed using the techniques previously described for two-part image coding.  Since wavelets are a general-purpose representation technique, they are also effective in representing the noise associated with spectral data.

To decompress audio, the lossless spectral signals are decompressed and summed with the decompressed lossy spectral noise functions.  A discrete inverse Fourier transform is applied to each resulting spectrum to produce approximations to the original time series of each block.  The blocks are then concatenated in their original ordering.  This results in an approximation to the original audio.

As a simple example of the superiority of a spectral representation for audio coding, we consider a trivial audio signal which is simply the interference between two sinusoids, 441Hz and 450Hz, each having amplitude 0.5.  The interference pattern repeats with a frequency of $450-441=9$ Hz, or about 3.26 times over a duration of approximately 0.307 seconds.  These divide the sampling frequency by 100 and 98, respectively, and the Fourier spectrum is correspondingly simple.  The graph of the spectrum has height one and width 13554, which is the number of samples in the clip.  It is solid black except for two pixels (100,1) and (98,1) which are at $50\%$ gray.

First, we will consider the time-domain representation.  The complexities at depths 0-15 are:
\begin{center}
\begin{tabular}{c}
66\\
2072\\
3754\\
4682\\
4177\\
3828\\
3627\\
3553\\
3506\\
3520\\
3529\\
3532\\
3532\\
3542\\
3544\\
3544
\end{tabular}
\end{center}

Note that in this case, higher bit depths are more redundant and lower bit depths are more complex, which inverts the behavior exhibited by pictures.  Also note that the total redundancy (7.64:1) is only a little over twice what would be expected from roughly 3.26 repetitions of any random signal.

Let's consider now the spectral case, which we have represented as the 13544x1 image described earlier, the spectrum downsampled to 8 bits.  Its compression performance for depths 0-8 is given below:

\begin{center}
\begin{tabular}{c}
122\\
132\\
132\\
132\\
132\\
132\\
132\\
132\\
132\\
\end{tabular}
\end{center}

In this case, compression performance is radically improved by transforming into frequency space.  For this purely sinusoidal construction, its 8-bit spectrum (which in this trivial case converts exactly to the 16-bit spectrum) compresses losslessly over 205:1, a vast improvement over compression in the time domain.  Because sine waves are eigenfunctions of the wave equation, two-part coding of an audio spectrum often outperforms the two-part coding of audio waveforms or time series.  Other phenomena sampled above their Nyquist frequency might benefit from a similar transformation.

\subsection{Example of Image Pattern Recognition}

Critical signals are useful for inference for several reasons.  On one hand, a critical signal has not experienced information loss - particularly, edges are preserved better since both the 'ringing' artifacts of non-ideal filters (the Gibbs phenomenon) and the influence of blocking effects are bounded by the noise floor.   On the other hand, greater representational economy, compared to other bit depths, translates into superior inference.

We will now evaluate the simultaneous compressibility of signals in order to produce a measure of their similarity or dissimilarity.  This will be accomplished using a sliding window which calculates the conditional prefix complexity $K(A|B)=K(AB)-K(B)$, as described in the earlier section regarding artificial intelligence.

In order to begin, we must first decide how to combine signals A and B into the simultaneous signal AB.  When processing textual data, the classic solution would be to concatenate the string representations of A and B.  With image data, other solutions are conceivably viable.  For instance, the images could be averaged or interlaced rather than simply concatenated.  However, numerical experiments reveal that simple concatenation results in superior compression performance.  This translates naturally into superior inference.

We will now calculate distance functions and convert them into pattern-specific filters for purposes of visualization.  The dimensions (x by y) of the texture B and the matrix representation of its distance function $D_{mn}(A,B)$ are used as input to construct a filter $f_{mn}$ selecting regions of image A that match pattern B.  In order to match the search space, the coordinates of the distance function associated with the sliding window is translated to the center of the window.  Each pixel in the filter assumes the value of one minus $D_{mn}(A,B)$, where $m+\lceil \frac{x}{2}\rceil,n+\lceil\frac{y}{2}\rceil$ is the center of the window closest to that pixel.  Since the distance function is not defined on the boundaries of the image, the closest available distance function may not be a good estimate, and distance estimates on the boundary should be regarded as being approximate, or omitted entirely.  They are shown in the examples for demonstrative purposes.

Multiplying pixels in the image A by the filter ($f_{mn}(A,B) A_{mn}$) has the effect of reproducing pixels that match the pattern while zeroing pixels that do not match.  Applying the filter multiple times (or exponentiating it) will retain the most similar regions and deemphasize less similar regions.  Alternately, one could apply a threshold to this function to produce a binary-valued filter denoting a pixel's membership or non-membership in the pattern set.

Using this visualization, we will see that the second order critical depth leads to better inference than the other bit depths.  If too much bit depth is used, there may be enough data to identify periodicity where it exits, overemphasizing matches similar to particular instances in the texture.  If bit depth is insufficient, many samples become equivalent to one another, leading to false positive matches.  This notion is similar to the considerations involved in 'binning' sparse data into histograms - if bins are too large, corresponding to insufficient bit depth, the histogram is too smooth and useful information is lost.  If the bins are too small, corresponding to excess bit depth, then there aren't enough statistics, leading to sample noise and significant error in the resulting statistics.  Ideally, we wish to have all the bit depth relevant to inference without including the superfluous data that tend to confound the inference process, and we will demonstrate that this occurs near the second-order critical depth, as one might expect, given its representational economy.

We will perform inference using the famous 'Lena' test image, which is 'lena3' among the Waterloo test images.  The image is 512x512 and has three channels of 8-bit color.  As a simple example of image recognition, we will will crop a sample texture from the lower 12 rows of the image, leaving a 512x500 image to generate our search space.  The search texture is taken from a rectangle extending between coordinates (85,500) and (159,511) in the original test image, using 0-based indexes, giving it dimensions of 75x12.  We wish to compare this texture (it samples the tassel attached to Lena's hat) against our search space, the set of rectangles in the cropped image, by estimating the conditional prefix complexity of the texture given the contents of each sliding window.

For reasons of computational expense, it may not be possible to evaluate complexity at every point in the search space.  For some applications, only the extremum is of interest, and nonlinear optimization techniques may be applied to search for this extremum.  In the case at hand, however, we wish to visualize the distance function over the entire search space, so we simply sample the distance function at regular intervals.  In this example, we will evaluate the distance function at points having even coordinates.

Though this exercise illustrates distance functions at each bit depth, which should make superior inference subjectively apparent, we also wish to estimate the second-order critical depth.  This will demonstrate that the signal which we expect \emph{a priori} to be the most economical also leads to superior inference.  We will start by estimating the complexity of the original 512x512 Lena image.  For bit depths $n=0-8$, $K(A^n)$ is:
\begin{center}
\begin{tabular}{cc}
  0 & 452\\
  1 & 23515\\
  2 & 43465 \\
  3 & 74632 \\
  4 & 118824 \\
  5 & 188461 \\
  6 & 284150 \\
  7 & 384018 \\
  8 & 482483
\end{tabular}
\end{center}

If we allow a $25\%$ tolerance factor within the maximum set, as before, we see that the first-order critical point is at depth 6 and the second order critical point is at depth 4.  We will see that inference at depth 4 outperforms inference at lower or higher bit depths, as predicted.

A sliding window is applied to the image to produce string B, and the conditional prefix complexity $K(A|B)=K(AB)-K(B)$ is calculated.  This is done for signals having bit depths of 1 through 8.  For visibility, the resulting filter is applied to the image four times and the resulting image is normalized to $100\%$ intensity.

The result follows.

\pagebreak

\begin{center}
  \includegraphics[width=3.4in]{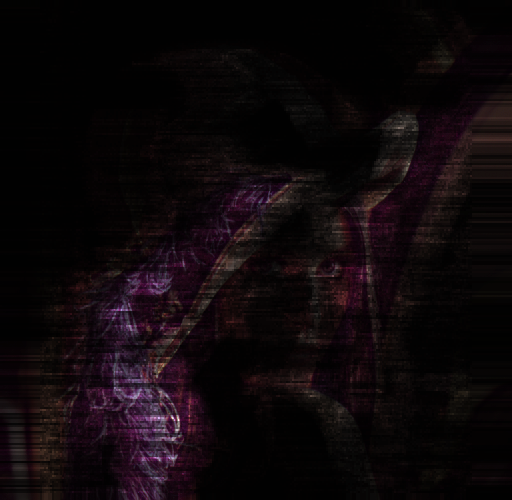}\\
\bigskip
  \includegraphics[width=3.4in]{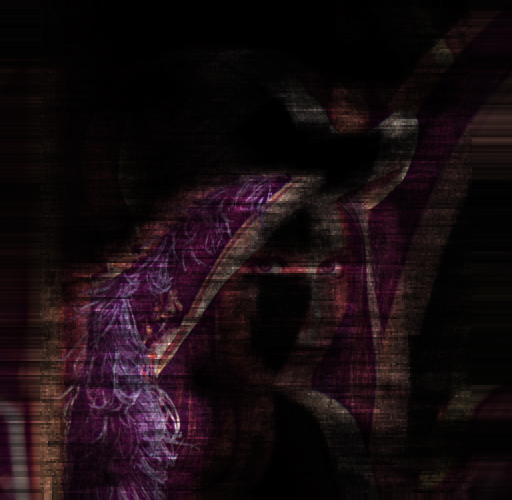}\\
  \smallskip
Depths 1 and 2.\\
  \includegraphics[width=3.4in]{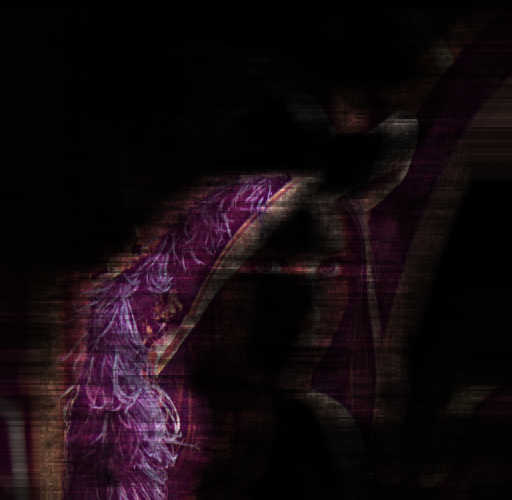}\\
\bigskip
  \includegraphics[width=3.4in]{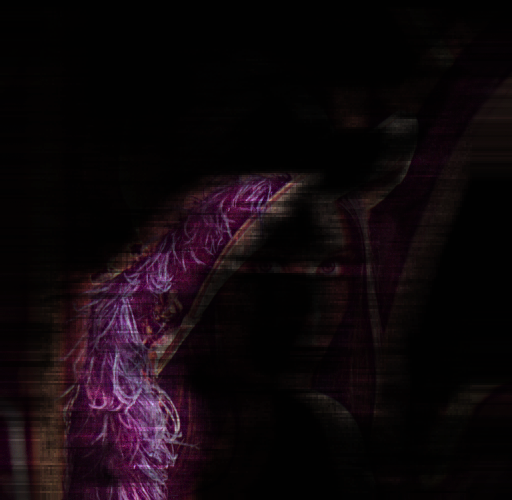}\\
\smallskip
Depths 3 and 4.\\
  \includegraphics[width=3.4in]{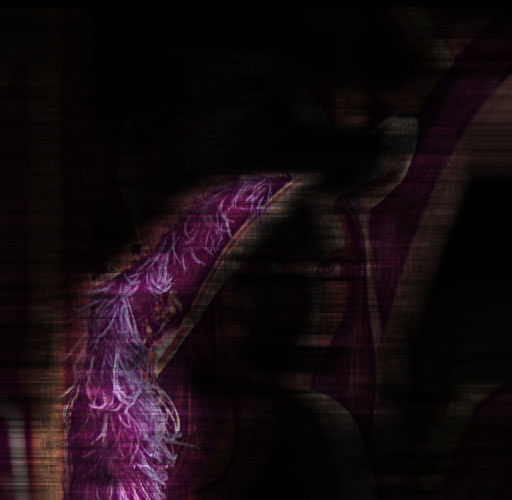}\\
\bigskip
  \includegraphics[width=3.4in]{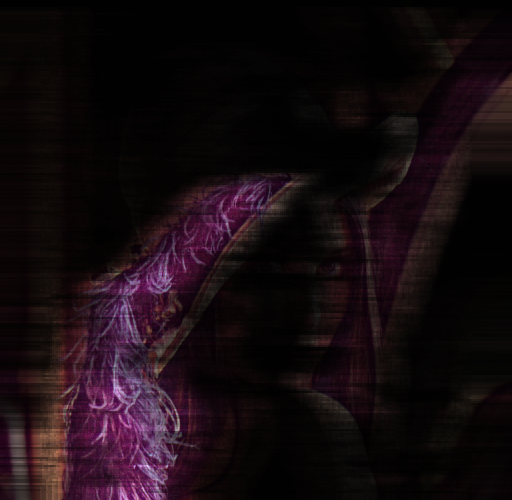}\\
\smallskip
Depths 5 and 6.\\
  \includegraphics[width=3.4in]{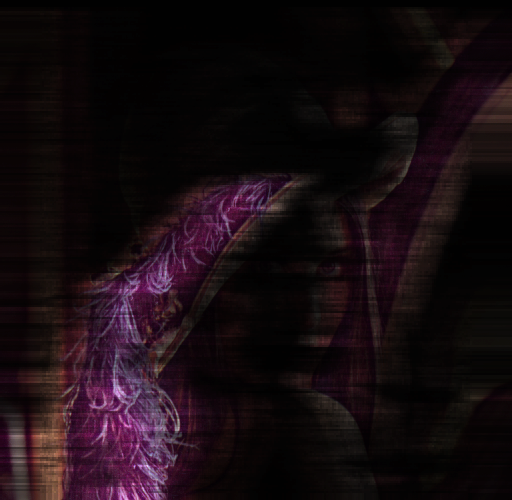}\\
\bigskip
  \includegraphics[width=3.4in]{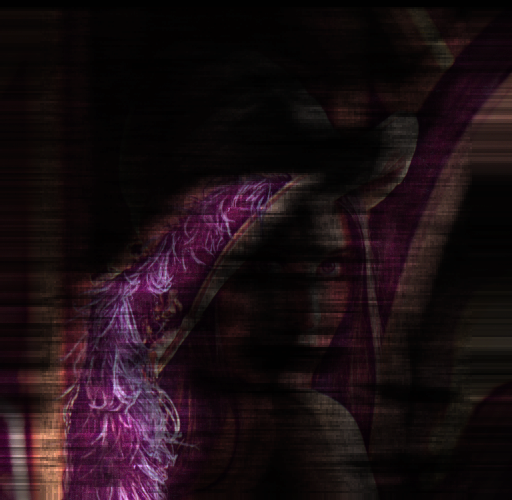}\\
\smallskip
Depths 7 and 8.\\
\end{center}
Figures 24-31: Pattern search, depths 1-8.  Depth 4 has fewer false positives than other depths as its greater economy translates into superior inference.  At lower depths, more false matches occur, since more of the image looks similar at this depth.  The full image at depth 8 has strong false matches and inferior performance even though it contains all available information, giving too much weight to bits which do not contain useful information about the signal.  This tends to overweight, for instance, short and possibly irrelevant literal matches.  Signals at depths 5-8 also exhibit this phenomenon to a lesser extent.

\section{Comments}

A critical depth (or other parameter, such as scale) represents 'critical' data in two senses of the word: on one hand, it measures the critical point of a phase transition between noise and smoothness, on the other, it also quantifies the essential information content of noisy data.  Such a point separates lossless signals from residual noise, which is compressed using lossy methods.
The basic theory of using such critical points to compress numeric data has now been developed.  This theory applies to arrays of any dimension, so it applies to audio, video, and images, as well as many other types of data.  Furthermore, we have demonstrated that this hybridization of lossless and lossy coding produces competitive compression performance for all types of image data tested.  Whereas lossy transformation standards such as JPEG2000 sometimes include options for separate lossless coding modes, a two-part code adapts to the data and smoothly transitions between the two types of codes.  In this way two-part codes are somewhat unique in being efficient for compressing both low-entropy and high-entropy sources.

The optional integration of Maximum Likelihood models and Monte-Carlo-type sampling is a significant departure from deterministic algorithms for data compression and decompression.  If sampling is employed, the decompression algorithm becomes stochastic and non-deterministic, potentially producing a different result each time decompression occurs.  The integration of statistical modeling into such an algorithm enables two-part codes which are engineered for specific applications.  This can lead to much higher levels of application-specific compression than can be achieved using general-purpose compression, as has been illustrated using a simple image corrupted by noise.

The test images presented use a bit depth of 8 bits per channel, as is standard in most of today's consumer display technology.  However, having more bits per sample (as in the proposed HDR image standard, for instance) means that the most significant bits represent a smaller fraction of the total data.  As such, the utility of a two-part code is increased at the higher bit depth, since more of the less-significant bits can be highly compressed using a lossy code, while the significant bits still use lossless compression.

Likewise, high-contrast applications will benefit from the edge-preserving nature of a two-part code.  Frequency-based methods suffer from the Gibbs' phenomenon, or 'ringing,' which tends to blur high-contrast edges.  In the approach described, this phenomenon is mitigated by limited use of such methods.  A two-part code should perform well in applications in which the fidelity of high-contrast regions is important.

As suggested previously, a two-part code can significantly outperform lossy transforms by many orders of magnitude for computer-generated artwork, cartoons, and most types of animation.  The low algorithmic complexity intrinsic to such sources leads to efficiently coded signals.

In most test cases, critical compression also outperforms JPEG2000 by orders of magnitude for black-and-white images.  Without the advantage of separating color data into chroma subspaces, the JPEG algorithms seem much less efficient.  For this reason, two-part codes seem to outperform JPEG coding in many monochrome applications.

JPEG2000 does perform well for its intended purpose - generating a highly compressed representation of color photographs.  For most of the color test photographs, a two-part code overtakes JPEG2000 at high quality levels.  The point at which this occurs (if it does) varies by photograph.  At relatively low bitrates, JPEG2000 usually outperforms a two-part code, but usually by less than an order of magnitude.  All examples presented to this point were directly coded in the RGB color space.  Since the theory of two-part codes applies to any array of n-bit integers, we could have just as easily performed analysis in the $Y C_b C_r$ color space, like the JPEG algorithms, which often improves the redundancy apparent in color data.  In the second part of the appendix, $RGB$-space critical compression will be compared to $Y C_b C_r$-space critical compression for color photographs.

One unique aspect of two-part data compression is its ability to code efficiently over a wide variety of data.  It can efficiently code both algorithmically generated regular signals and stochastic signals from empirical data.  The former tends to be periodic, and the random aspects of the latter tend to exhibit varying degrees of quasiperiodicity or chaos.  However, the creation of periodicity or redundancy is essential to the comparison operation - the prefix complexity involves concatenation, which becomes similar to repetition if the concatenated objects are similar.  Concatenation can create periodicity from similarities, even if the objects being concatenated have no significant periodicity within themselves, as may be the case with data altered by noise.  The inferential power of critical compression derives from its ability to compress periodicity which would otherwise be obscured by noise.

In spite of its ultimately incalculable theoretical underpinnings, the human eye intuitively recognizes a critical bit depth from a set of truncated images.  The mind's eye intuitively recognizes the difference between noisy, photographic, "real world" signals and smooth, cartoon-like, artificial ones.  Human visual intelligence can also identify the effective depth from the noisy bits - it is the depth beyond which features of the original image can be discerned in the noise function.  Conversely, given a computer to calculate the critical point of an image, we can determine its critical information content.  Since noise can't be coded efficiently due to its entropy, an effective learner, human or otherwise, will tend to preferentially encode the critical content.  This leads directly to more robust artificial intelligence systems which encode complex signals in a manner more appropriate for learning.

\section{Acknowledgements}

This work was funded entirely by the author, who would like to acknowledge his sister, Elizabeth Scoville, and his parents, John and Lawana Scoville.  A patent related to this work is pending.

\providecommand{\bysame}{\leavevmode\hbox to3em{\hrulefill}\thinspace}
\providecommand{\MR}{\relax\ifhmode\unskip\space\fi MR }
\providecommand{\MRhref}[2]{%
  \href{http://www.ams.org/mathscinet-getitem?mr=#1}{#2}
}
\providecommand{\href}[2]{#2}

\pagebreak

\section{Appendix: Image Compression Performance}

In the following plots, each solid line represents two-part codes having various lossy bitrates at a particular signal depth.  The dotted lines show the error level at various bitrates of JPEG2000 coding, these are also two-part codes at signal depth zero.

\subsection{University of Waterloo Test Images}

The image repository maintained by the University of Waterloo's fractal coding and analysis group contains 32 test images.  The collection includes a wide variety of content, with photographic and computer generated content in both color and black and white.  Two part coding dominates direct lossy image coding for the majority of these images, demonstrating the power and versatility of critical data compression using two-part codes.  The images which perform better with direct lossy coding are generally color photographs, with JPEG2000 having its greatest advantage at low quality levels.  Two-part codes seem to have an advantage for the other images, sometimes by multiple orders of magnitude.

\pagebreak

\begin{center}
% GNUPLOT: LaTeX picture
\setlength{\unitlength}{0.240900pt}
\ifx\plotpoint\undefined\newsavebox{\plotpoint}\fi
\sbox{\plotpoint}{\rule[-0.200pt]{0.400pt}{0.400pt}}%

\end{center}

\pagebreak

\subsection{Kodak Photo CD Test Images}

The method described was applied to 24 uncompressed 24-bit photographic images from a sample Kodak Photo CD.  We compare critical compression at various bitdepths in the $RGB$ color space using PAQ8l and JPEG2000, as before, against a YCbCr-space encoding which critically compresses a luma ($Y$) channel at various bitdepths using PAQ8l and the chroma channels ($C_b$ and $C_r$) using JPEG2000.  For this transformation, the chroma parameters $k_b$ and $k_r$ are both equal to $\frac{1}{3}$, making the $Y$ channel a simple average of the corresponding red, green, and blue color values.  The results (with $RGB$ above and $YC_{r}C_b$ below) show that while JPEG2000 retains the advantage at low to moderate quality levels, the critical luma/lossy chroma $YC_{r}C_b$ scheme is usually more efficient at moderate to high quality levels than direct JPEG2000 coding or critical compression in $RGB$.

\pagebreak

\begin{center}
% GNUPLOT: LaTeX picture
\setlength{\unitlength}{0.240900pt}
\ifx\plotpoint\undefined\newsavebox{\plotpoint}\fi
\sbox{\plotpoint}{\rule[-0.200pt]{0.400pt}{0.400pt}}%


\end{center}


\begin{thebibliography}{10}

\bibitem{AT05}
T.~Acharya and P.~Tsai, \emph{Jpeg2000 standard for image compression}, Wiley,
  Hoboken, NJ, 2005.

\bibitem{CH87}
G.J. Chaitin, \emph{Algorithmic information theory}, Cambridge University
  Press, 1987.

\bibitem{CO91}
T.M. Cover, \emph{Elements of information theory}, Wiley, 1991.

\bibitem{FI21}
R.A. Fisher, \emph{On the mathematical foundations of theoretical statistics},
  Phil. Trans. of the Royal Society of London \textbf{Series A 222} (1921),
  309--368.

\bibitem{GL96}
M.~Gell-Mann and S.~Lloyd, \emph{Information measures, effective complexity,
  and total information}, Complexity \textbf{2/1} (1996), 44--52.

\bibitem{HA28}
R.~V.~L. Hartley, \emph{Transmission of information}, Bell System Technical
  Journal (July 1928), 535--?

\bibitem{Jackson}
J.~Jackson, \emph{Classical electrodynamics}, third ed., John Wiley and Sons,
  New York, 1999.

\bibitem{Knuth2}
D.~Knuth, \emph{The art of computer programming, vol 2.}, Addison-Wesley,
  Reading, MA, 1998.

\bibitem{LV97}
M.~Li and P.~Vit\'{a}nyi, \emph{An introduction to kolmolgorov complexity and
  its applications}, second ed., Springer-Verlag, New York, 1997.

\bibitem{MT00}
G.~Marsaglia and W.W. Tsang, \emph{The ziggurat method for generating random
  variables}, Journal of statistical software \textbf{5} (2000).

\bibitem{ny28}
H.~Nyquist, \emph{Certain topics in telegraph transmission theory}, Trans. AIEE
  \textbf{47} (1928), 617--644.

\bibitem{me1}
J.~Scoville, \emph{On macroscopic complexity and perceptual coding},
  arXiv:1005.1684.

\bibitem{SH48}
C.E. Shannon, \emph{The mathematical theory of communication.}, Bell Labs Tech.
  J. \textbf{27} (1948), 379--423,623--656.

\bibitem{sh49}
\bysame, \emph{Communication in the presence of noise}, Proc. Institute of
  Radio Engineers \textbf{37} (1949), 10--21.

\bibitem{ZU89}
W.H. Zurek, \emph{Algorithmic randomness and physical entropy.}, Physical
  Review, Ser. A \textbf{40(8)} (1989), 4731--4751.

\end{thebibliography}
\end{document}